\documentclass[12pt]{iopart}
\usepackage[utf8]{inputenc}
\usepackage{graphicx,color}
\usepackage{caption}
\usepackage{iopams}
\usepackage{epstopdf}
\usepackage{soul}
\usepackage{adjustbox}
\usepackage{tabularx}
\newcommand{\D}{\mathrm{d}}
\newcommand{\half}{\frac{1}{2}}

\newcommand{\be}{\begin{equation}}
\newcommand{\ee}{\end{equation}}
\newcommand{\bea}{\begin{eqnarray}}
\newcommand{\eea}{\end{eqnarray}}
\newcommand{\ba} {\begin{eqnarray} }
\newcommand{\ea} {\end{eqnarray} }
\newcommand{\eps}{\epsilon}
\newcommand{\s}{\mathrm{c}}
\newcommand{\n}{\mathrm{m}}
\newcommand{\re}{\mathrm{Re}}

\newcommand{\vecr}{\bi{r}}
\newcommand{\vecq}{\bi{q}}
\newcommand{\vecu}{\bi{u}}
\newcommand{\vecp}{\bi{p}}
\newcommand{\vecv}{\bi{v}}
\newcommand{\vecj}{\bi{j}}
\newcommand{\vech}{\bi{h}}
\newcommand{\vecx}{\bi{x}}
\newcommand{\vecf}{\bi{f}}
\newcommand{\vecfrel}{\bi{f}^{\rm rel}}
\newcommand{\tenQ}{\bi{Q}}
\newcommand{\teneps}{\bepsilon}

\newcommand{\kbt}{k_{\mathrm{B}}T}

\begin{document}


\title{Active-gel Theory for Multicellular Migration of Polar Cells in the Extra-cellular Matrix}

\author{Ram M. Adar$^{1,2,3}$ and Jean-François Joanny$^{1,2,3}$}
\address{$^1$ Collège de France, 11 place Marcelin Berthelot, 75005 Paris, France\\
$^2$ Laboratoire Physico-Chimie Curie, Institut Curie, Centre de Recherche, Paris Sciences et Lettres Research University, Centre National de la Recherche Scientifique, 75005 Paris, France \\ $^3$ Université Pierre et Marie Curie, Sorbonne Universités, 75248 Paris, France
}


\begin{abstract}
    We formulate an active-gel theory for multicellular migration in the extra-cellular matrix (ECM). The cells are modeled as an active, polar solvent, and the ECM as a viscoelastic solid. Our theory enables to analyze the dynamic reciprocity between the migrating cells and their environment in terms of distinct relative forces and alignment mechanisms. We analyze the linear stability of polar cells migrating homogeneously in the ECM. Our theory predicts that, as a consequence of cell-matrix alignment, contractile cells migrate homogeneously for small wave vectors, while sufficiently extensile cells migrate in domains. Homogeneous cell migration of both extensile and contractile cells may be unstable for larger wave vectors, due to active forces and the alignment of cells with their concentration gradient. These mechanisms are stabilized by cellular alignment to the migration flow and matrix stiffness. They are expected to be suppressed entirely for rigid matrices with elastic moduli of order $10$\,kPa. Our theory should be useful in analyzing multicellular migration and ECM patterning at the mesoscopic scale.

\end{abstract}

\maketitle



\section{Introduction}
\label{sec1}
Multicellular migration plays a key role during development, wound healing and metastasis~\cite{Hakim2017,Alert2019,Friedl2009}. A basic distinction can be made between solid-like and fluid-like migration, which differ in the strength and duration of cell-cell adhesions. Fluid-like migration is referred to as ``multicellular streaming''~\cite{Friedl2012,Clark2015} and is the main motivation for this paper.
The migration mode depends on the properties of the cells and their environment. Polarization is important for cell migration in both the single-cell and multicellular levels. Intuitively, cells with a well-defined direction migrate in this direction. The constant crosstalk between migrating cells and their environment is also gaining increasing attention as an essential factor for multicellular migration~\cite{Clark2015,Charras2014,Alexander2016,VanHelvert2018}. This  is referred to as ``dynamic reciprocity''~\cite{Alexander2016} or ``mechanoreciprocity''~\cite{VanHelvert2018}.

We focus on migration that takes place in the extra-cellular matrix (ECM), which consists mostly of collagen I. It was shown that anisotropic ECM organization with aligned collagen fibres promote cancer-cell migration in collagen tracks~\cite{Clark2015,Han2016}. Cells are able to remodel the fibres and change their environment, either mechanically or chemically. For example, interactions between ECM fibers and fibroblasts tune ECM properties and can account for variations in matrix isotropy, density, and homogeneity found across different tissues~\cite{Wershof2019}. The viscoelastic nature of the matrix is also important in  tumor growth and cancer-cell invasion~\cite{Elosegui-Artola2021}. For example, collagen relaxation was shown to drive the motion of  cancer-cell clusters on collagen gels in two dimensions~\cite{Clark2020}.

While several mathematical models have been proposed for multicellular migration in the ECM in different contexts~\cite{Murray1983,Oster1983,Olsena1997,Dallon1999,McDougall2006,Painter2009}, a physical understanding of cell-ECM interaction at the mesoscopic scale and in three dimensions is still missing. Here, we propose to describe the ECM together with the migrating cells as an active, permeating, polar gel. Such systems were studied in the past in different contexts~\cite{Joanny2007, Callan-Jones2011,Callan-Jones2013,Pleiner2013,Pleiner2016,Maitra2018,Adar2021}. We rely mostly on our recent work~\cite{Adar2021}, which explored permeation instabilities of an active, polar, solvent immersed in a viscoelastic fluid. This theory was formulated as a two-fluid model, with a clear distinction between forces that act on the network and solvent separately and relative forces between them.

In this work, we develop our theory further and adapt it to cells in the ECM. We formulate a solid-fluid model, considering that the ECM is solid at long times, and take into account cell division and strain-polarization alignment. We analyze the linear stability of a homogeneous flow of polarized cells. Instabilities infer transient and possibly long-lived, migrating cell collections.

Our key findings are: 1) Cell-matrix interactions can be classified according to their characteristic spatial order, dynamics, activity, reversibility, and elasticity. 2) Active stresses can destabilize flexible matrices.  3) Matrix stiffness stabilizes the ECM and suppresses alignment-driven instabilities. 4) ECM stability for small wave vectors is determined by the active nematic stress; it is stable for contractile cells and unstable for sufficiently extensile cells. 5) Alignment of polarization to concentration gradients can either stabilize or destabilize the ECM, while alignment to the migration current stabilizes it. The former can change the transient domain size by orders of magnitude.

The outline of the paper is as follows: In Sec.~\ref{sec2}, we derive our theory for multicellular migration in the ECM, in terms of an active, polar fluid permeating in a viscoelastic solid. Next, we highlight in Sec.~\ref{sec3} the different matrix-cell interactions that arise naturally from out theory. In Sec.~\ref{sec4} we derive the linearized equations that determine the linear stability of the system. The analysis is performed in the isotropic case and in the rigid-matrix limit in Secs.~\ref{sec5} and \ref{sec6}, respectively. We analyze the stability in the general case in Sec.~\ref{sec7} and clarify the stabilizing or destabilizing role of cell-matrix alignment mechanisms. We conclude in Sec.~\ref{sec8} by discussing possible extensions of our theory and how it relates to biologically-relevant scenarios.

\section{Theory}
\label{sec2}
We consider a two-component gel, composed of active, polar cells (c) and a viscoelastic
matrix (m). The polarization field is given by the unit vector $\vecp$. The matrix is modeled as a viscoelastic solid; fluid at short times and solid at long times. Its deviation from the reference state is given by the displacement vector $\vecu$. The matrix has a volume fraction $\phi$, and the cells $1-\phi$. The gel is assumed to be incompressible.

The free energy of the gel can be written as
\ba
\label{eq1}
\fl F&=\int \D^3r\left[\kbt a^{-3}\left(1-\phi\right)\ln\left(1-\phi\right)+\phi\left(1-\phi\right)\left[\chi_0+\psi\, {\rm Tr} \left(\tenQ\teneps\right) \right]+\kappa\left(\nabla\phi\right)^2\right.\nonumber\\
\fl &\left.+K \left(\half\left(\nabla\vecp\right)^2-l_p^{-1}\vecp\cdot\nabla\phi\right)-\half h_\parallel \vecp^2+\frac{\phi}{\phi_0}\left(G\Tr\left(\tilde{\teneps}^2\right)+\half B\epsilon^2\right)\right].
\end{eqnarray}
The first line is the Flory-Huggins free energy of a binary mixture in the limit long polymer chains, where $k_B T$ is the thermal energy and $a$ is a microscopic length. The $\chi_0$ term  accounts for short-range interactions, and $\psi$ to an aligning interaction (see, e.g., \cite{Maitra2018,Warner2007liquid,Hemingway2015,Hemingway2016}) that is related to three-dimensional ``contact guidance'' in biological contexts. It is written in terms of the nematic tensor $Q_{\alpha\beta}=p_\alpha p_\beta-1/3\,\delta_{\alpha\beta}$ and linearized strain tensor $\epsilon_{\alpha\beta}=\left(\partial_\alpha u_\beta+\partial_\beta u_\alpha\right)/2$. It  results in a shear elastic stress in the reference state, which aligns the matrix parallel or normal to the polarization axis ($\psi<0$ or $\psi>0$, respectively). The $\kappa$ term accounts for the interfacial tension that suppresses large concentration gradients.

 The first part of the second line accounts for variations of the polarization field from the homogeneously polarized state~\cite{Kruse2005,Voituriez2006}, where $K$ is the Frank constant in the one-constant approximation. It is generally a function of the cellular volume fraction, but this dependence does not play any role in our linear analysis and is disregarded hereafter. The second term accounts for alignment with respect to concentration gradients in terms of the length $l_p$. This describes, for example, cellular alignment at cluster interfaces, similarly to anchoring at droplet interfaces~\cite{Cates2018}. We refer hereafter to this mechanism as ``concentration alignment''. The $h_\parallel$ term is a Lagrange multiplier to ensure that $\vecp^2=1$.

 The final contribution is the elastic free energy. For simplicity, we restrict ourselves to linear elasticity, where $G$ and $B$ are the shear and bulk moduli for $\phi=\phi_0$, respectively. We decompose the strain into the scalar $\epsilon=\epsilon_{\alpha\alpha}$ and the traceless tensor, $\tilde{\epsilon}_{\alpha\beta}=\epsilon_{\alpha\beta}-\epsilon/3\,\delta_{\alpha\beta}$.

We describe the dynamics of the concentration, polarization, and displacement fields within a thermodynamic
framework. The matrix moves with a velocity $\vecv^\n=\partial \vecu/\partial t$ and the cells with a velocity $\vecv^\s$, corresponding to a center-of-mass (COM) velocity, $\vecv=\phi\,\vecv^\n+\left(1-\phi\right)\vecv^\s$, and a relative current, $\vecj=\phi\left(1-\phi\right)
\left(\vecv^\n-\vecv^\s\right)$. We assume the same specific mass for both components.

Living cells are active. They are constantly driven out of equilibrium by the input of an energy $\Delta\mu$ that corresponds, for example, to the chemical potential difference between ATP and its hydrolysis products~\cite{Prost2015,Joanny2007}. In particular, the cells divide and die, while the mass of the matrix is conserved, i.e.,
\ba
\label{eq2}
\fl \partial_t \phi+\nabla\cdot\left(\phi\vecv^\n\right)=0,&
\qquad \partial_t\left(1-\phi\right)+\nabla\cdot\left[\left(1-\phi\right)\vecv^\s\right]&=\left(1-\phi\right) k\approx k_\phi\left(\phi-\phi_0\right).
\end{eqnarray}
The cellular growth rate, $k$, is generally a function of the pressure ~\cite{Basan2009,Ranft2010}. We linearize the right hand side around the homeostatic pressure and volume fraction $\phi_0$, where cell division and death balance each other, in terms of the rate $k_\phi>0$. In Eq.~(\ref{eq2}), cell division and death are related only to cell component.
A more detailed description would include a third solvent component that exchanges mass with the cells as part of these processes. Coarse-graining over the solvent neglects cell-solvent friction. This is reasonable because the cells are much more viscous than the solvent, making cell-matrix friction more important.
 This description also neglects active matrix deposition and degradation by the cells, which can be especially important for fibroblasts. The study of this effect is reserved for future work. Note that the incompressibility condition is affected by the active growth rate and is given by $\nabla\cdot\vecv=k_\phi\left(\phi-\phi_0\right)$.

For the polarization, we derive in~\ref{AppA} the following constitutive relation:
\begin{equation}
\label{eq3}
\fl \left(\partial_t+\vecv^\s\cdot\nabla\right)\vecp =\bar{h}_\parallel\vecp+\vecp\cdot\nabla\vecv^\s+D_p \nabla^2\vecp +D_p l_p^{-1}\nabla \phi+\lambda\vecj+\phi\left(1-\phi\right)\bar{\psi} \teneps\cdot\vecp.
\end{equation}
The first term on the right-hand side originates from $h_\parallel$ of Eq.~(\ref{eq1}) and ensures that $\vecp^2=1$.
 The next term accounts for the shear alignment of the cells as if they were solid rods (shear-alignment parameter of $-1$) and $D_p=K/\gamma_1$ is the angular diffusion constant, with $\gamma_1$ the rotational viscosity. The final three terms in Eq.~(\ref{eq3}) describe polarization alignment due to cell-matrix interaction. The first describes alignment to concentration gradients, while  $\lambda$ is the permeation-alignment constant~\cite{Adar2021} that describes how the cells align to their migration current in the matrix. Finally, $\bar{\psi}$ describes cell-matrix alignment due to the $\psi$ term in Eq.~(\ref{eq1}) and possible active mechanisms. We focus on passive alignment, for which $\bar{\psi}=-2\psi/\gamma_1$ (see~\ref{AppA}). These alignment mechanisms are illustrated in Fig.~\ref{fig1}.

Equations~(\ref{eq2}) and (\ref{eq3}) describe the dynamics of the cell concentration and orientation. They depend on the cell velocity and matrix displacement, which can be determined from force balance equations. We make use of a solid-fluid model, similar to the fluid-fluid model of Ref.~\cite{Adar2021}. Force-balance equations are written separately for the matrix and cells as
\ba
\label{eq4}
\vecf^{\n}-\phi\nabla\delta P =\vecfrel, &\qquad \vecf^{\s}-\left(1-\phi\right)\nabla\delta P=-\vecfrel,
\end{eqnarray}
where $\vecf^{\n}$ and $\vecf^{\s}$ are the forces acting on the matrix and cells, respectively, $\delta P$ is a pressure difference that enforces global incompressibility, and $\vecfrel$ is the relative force between the two components.

The forces acting on each of the components are
\ba
f^{\n}_\alpha&=\partial_\beta\left[\sigma^{\rm el}_{\alpha\beta}+ \frac{\phi}{\phi_0}\left(2G\tau\partial_t\tilde{\epsilon}_{\alpha\beta}+B\bar{\tau}\partial_t\epsilon\delta_{\alpha\beta}\right)\right]-\phi\partial_\alpha\bar{\mu},\label{eq5} \\
f^{\s}_\alpha&=\partial_\beta\left[-h_\alpha p_\beta+\left(1-\phi\right) \left(\bar{\zeta}\delta_{\alpha\beta}+\zeta Q_{\alpha\beta}\right)\right]-h_\beta\partial_\alpha p_\beta\label{eq6},
\end{eqnarray}
where the summation convention was used. Here, $\sigma^{\rm el}_{\alpha\beta}=\delta F/\delta \epsilon_{\alpha\beta}$ is the elastic stress and the next two terms describe the viscoelastic shear and compressional stresses, in terms of the shear and compressional retardation times, $\tau$ and $\bar{\tau}$, respectively. The last term is the osmotic pressure gradient, with the relative chemical potential $\bar{\mu}=\delta F/\delta\phi$.

In Eq.~(\ref{eq6}), the first term is the stress due to shear alignment, where $\vech=-\delta F/\delta \vecp$ is the orientational field. Next is the active cellular stress that consists of an isotropic contribution $\sim \bar{\zeta}$ and a traceless contribution $\sim\zeta$, proportional to the nematic tensor, $\tenQ$. The stresses $\bar{\zeta}$ and $\zeta$ are considered as constants, neglecting the possible dependence on matrix properties~\cite{De2007}.
The last term in Eq.~(\ref{eq6}) originates in the Ericksen stress and vanishes to linear order around a polarized state. The cellular viscous dissipation has been neglected; it is negligible compared to the relative friction force on lengthscales larger than the matrix mesh size, which are relevant to our hydrodynamic framework.

The cell-matrix relative force is given by
\be
\label{eq7}
\vecfrel=\frac{1}{\gamma}\vecj-\phi\left(1-\phi\right)\left(\lambda\vech+\nu\vecp+\nu'\teneps\cdot\vecp\right).
\ee
The first term is the friction force, where $\gamma$ is the mobility. The term $\sim\lambda$ is the reactive force associated with permeation alignment~\cite{Adar2021}. The last two terms are active forces that are the main contributors to cell motility in the polar case. The $\nu'$ term can also be related to anisotropic friction due to matrix strain, as is explained in~\ref{AppB}.

\section{Description of cell-matrix interaction}
\label{sec3}
\begin{figure}[ht]
\centering
\includegraphics[width=0.8\textwidth]{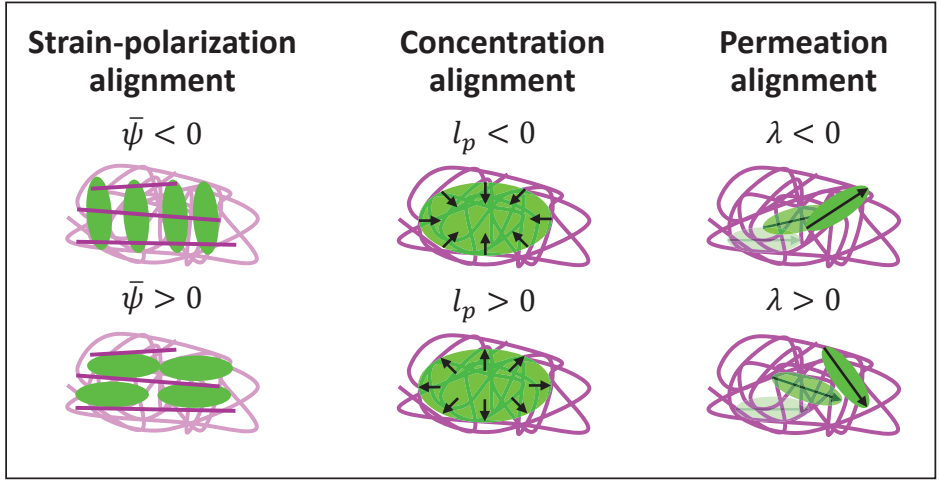}
\caption{(Color online) Heuristic description of alignment mechanisms of polar cells (green, polarization indicated by a black arrow) in a viscoelastic matrix (purple).  The strain-polarization term aligns the matrix segments normal or parallel to the polarization axis ($\bar{\psi}<0$ or $\bar{\psi}>0$, respectively). The concentration-alignment term aligns the polarization towards or away from gradients in cell concentration ($l_p<0$ or $l_p>0$, respectively). The permeation-alignment term aligns the cell towards or away from its direction of relative motion ($\lambda<0$ or $\lambda>0$, respectively).}
\label{fig1}
\end{figure}

 Our framework is convenient for analyzing the crosstalk between cells and their environment and classifying its underlying mechanisms.  First, the matrix and cells influence each other indirectly, because they are constrained by global force balance [sum of the two lines in Eq.~(\ref{eq4})]. At the same time, each component undergoes convection according to its own velocity, which is also determined from the force-balance equations. More interestingly, we can identify and classify mechanisms of direct interaction between the cells and the matrix. These include the friction force and active relative forces, as well as permeation alignment, concentration alignment, and strain-polarization alignment. The three alignment mechanisms are illustrated in Fig.~\ref{fig1}.

The interaction terms require a combination of matrix and cells and vanish in pure phases $\left(\phi=0,1\right)$. They are especially important in the case of multicellular streaming and small matrix mesh size, where cells can flow and mix with the matrix on a mesoscopic scale. This mixing yields bulk interaction terms and relative forces that exist between the cells  and matrix within each volume element, rather than surface terms that exist only between clearly separated phases.

\begin{table}[ht]
\caption{Classification of cell-matrix interaction terms.}
\begin{adjustbox}{width=1\textwidth}
\begin{tabular}{| l | c | c | c | c | c |}
   \mr
   \bf Mechanism& \bf Dynamics & \bf Reversibility & \bf Activity & \bf Symmetry & \bf Elasticity \\ \hline
Friction & dynamic & dissipative & passive & isotropic & none \\ \hline
Active relative force & dynamic & reactive & active & polar & none/elastic\\ \hline
Permeation alignment & dynamic & reactive & passive & polar & none\\ \hline
Concentration alignment & static & dissipative & passive & polar & none\\ \hline
Strain-polarization alignment & static & dissipative & passive / active & nematic & elastic \\ \hline
  \end{tabular}
  \end{adjustbox}
  \label{table1}
 \end{table}

The thermodynamic framework allows to classify these mechanisms according to five categories (and see Table~\ref{table1}); {\it Dynamics} - requires cell migration (dynamic) or not (static). {\it Reversibility} - produces entropy (dissipative) or not (reactive). {\it Activity} - requires ATP hydrolysis (active) or not (passive). {Symmetry} - requires polar order, nematic order, or no order (isotropic). {Elasticity} - requires elasticity or not. As is evident from Table~\ref{table1}, each mechanism is unique according to this classification. This demonstrates that the different terms of our theory have distinguishable properties and can be inferred from sufficient experimental data.
\section{Linear stability analysis}
\label{sec4}
The system has a homogeneous steady state at the homeostatic concentration $\phi=\phi_0$ and is in a fully polarized state that we set as $\vecp^0=\hat{\bf{x}}$. The active relative force drives a homogeneous steady-state current $\vecj^0=j_0\vecp^0$ (see ~\ref{AppB}). This effect is purely active and polar.
As the concentration and polarization are homogeneous, the active cellular stress and matrix alignment stress are constant, and the matrix displacement is determined by the boundary conditions. We consider the case where the matrix is in its equilibrium configuration, meaning that the stress on the boundaries matches the active stress. The steady-state strain is then given by $\teneps=-\psi\phi_0\left(1-\phi_0\right)\tenQ/2G$.
As cells are mostly known to align parallel to matrix segments, the sign $\psi<0$ is chosen.

We analyze the linear stability of the steady state with respect to perturbations with a growth rate $s$ and wave vector $q$, of the form $\vecx=\vecx^0+\vecx^1\exp\left(st+i\vecq\cdot\vecr\right)$, where $\vecx=\left(\phi,\vecp,\vecu\right)$. For simplicity, we focus on wave vectors perpendicular to the steady-state polarization, $q_x=0$, assuming that heterogeneity is most notable normal to the direction of migration.
As the matrix is elastic, its concentration changes only via strain, according to $\phi^1/\phi_0=-\eps^1$. This relates the normal components of the displacement, $u^1_y$ and $u^1_z$ to $\phi^1$. In addition, $\phi^1$ is affected only by the divergence of the polarization, $i \vecq\cdot\vecp^1\equiv p^1_d$. This is reasonable because, around the polarized state and to linear order, this is the only scalar obtained from $\vecp^1$.
These arguments reduce the dimensions of the linear stability analysis to three, corresponding to $\phi^1$, $p_d^1$, and $u^1_x$.

We find $u^1_x$ as a function of the concentration and polarization and obtain the following linearized equations (see ~\ref{AppB}):
\ba
 s\phi^1&=-\left[k_\phi+\left(D_\phi+l_\eta^2 s\right)q^2\right]\phi^1-\left[j_0+j_u u_p+D_p\left(l_p^{-1}-\lambda\right)l_{\gamma_1}^2q^2\right]p_d^1,\label{eq8}\\
 s p^1_d&=-\left[\left(\bar{\psi}-\lambda j_u\right)u_p+D_p\left(1+\lambda\left(\lambda-l_p^{-1}\right)l_{\gamma_1}^2\right)q^2\right]p^1_d\nonumber\\&+\left[\lambda\left(D_\phi+l_\eta ^2 s\right)-l_p^{-1}D_p\right]q^2\phi^1.\label{eq9}
\end{eqnarray}
The parameters that appear in Eqs.~(\ref{eq8}) and (\ref{eq9}) are listed in Table~\ref{table2}. In Eq.~(\ref{eq8}), the first term describes the concentration relaxation due to active cell division and death. The terms quadratic in $q$ account for osmotic diffusion, where $D_\phi$ is the effective diffusion coefficient in the presence of elasticity, permeation alignment and active cellular stress. $l_\eta=\sqrt{\left(1-\phi_0\right)\gamma\left(4G\tau/3+B\bar{\tau}\right)/\phi_0}$ is a screening length that arises from the interplay between transient matrix viscosity and cell-matrix friction.

The second part of Eq.~(\ref{eq8}) accounts for the relative force in the direction of $\vecp^1$. The first two terms relate to the active relative current. $j_0$ is the steady-state current, while $j_u=\gamma\left(2\lambda\phi_0\left(1-\phi_0\right)\psi-\nu'\right)$ describes a correction due to the network strain. The term $u_p$ is a function of the rate $\tau s$ and is related to the network displacement in the $x$-direction, due to strain-polarization coupling (see Sec.~\ref{sec6} and ~\ref{AppB}). The relative force quadratic in $q$ originates from concentration-polarization alignment, with $l_{\gamma_1}=\sqrt{\phi_0\left(1-\phi_0\right)\gamma\gamma_1}$ being a screening length due to the interplay between rotational viscosity and friction.

In Eq.~(\ref{eq9}), the first term describes a $q^0$ polarization rate resulting from network-strain coupling (see Sec.~\ref{ssec71}). The term quadratic in $q$ is the effective angular diffusion constant. Alignment to concentration gradients and flow may render it negative. The second line of Eq.~(\ref{eq9}) describes the two mechanisms of polarization rotation due to concentration gradients; one is dynamic (permeation alignment $\sim\lambda$) and the second is static (concentration alignment $\sim l_p^{-1}$). The two mechanisms either add up or compete with each other.

This linear set of equation can be written as $M\cdot \vecx=0$, where $\vecx=\left(\phi^{1},p_{d}^{1}\right)^{T}$. The dispersion relation $s\left(q\right)$ is found by solving det$M=0$. The system is stable if $\re\, s<0$ for all the eigenvalues of the linear system. The stability analysis is involved, due to the large number of mechanisms that take place. Therefore, we focus first on two limiting cases of isotropic cells and a rigid matrix. Then, we analyze separately the different alignment mechanisms and their effect on stability.

\begin{table}[ht]
	\caption{Parameters of the theory. Estimations of the parameters are found in~\ref{AppD}.}
	\begin{adjustbox}{width=1\textwidth}
		\begin{tabular}{| c | l |c | l |}
			\mr
			\bf Symbol& \bf Description & \bf Symbol & \bf Decription \\ \hline
			$D_\phi$ & effective osmotic diffusion constant & $l_\phi$ & interfacial correlation length \\ \hline
			$k_\phi$ & cellular division rate & $D_p$ & angular diffusion constant  \\ \hline
			$l_\eta$ & screening length due to matrix viscosity & $l_{\gamma1}$ & screening length due to rotational viscosity\\ \hline
			$\bar{\psi}$	& strain-polarization alignment rate & $u_p$ & measure of polarization-induced network $x$-displacement \\ \hline
			$j_0$ & steady-state relative current & $j_u$ & $j_u u_p$ is the strain-induced relative current \\ \hline
			$l_p$ & concentration-alignment coupling (length) & $\lambda$ & permeation-alignment coupling (inverse length)\\ \hline
			
		\end{tabular}
	\end{adjustbox}
	\label{table2}
\end{table}

\section{ Isotropic case: active stresses may destabilize a flexible ECM.}
\label{sec5}

Cells are in many cases isotropic. The polarization terms then drop out of the equations, and the dispersion relation is found from Eq.~(\ref{eq8}) as  $s=-\left(k_\phi+D_\phi q^2\right)/\left(1+l_\eta^2q^2\right)$. This situation was explored by Murray, Oster, and Harris~\cite{Murray1983,Oster1983} in their works on  mesenchymal morphogenesis, which similarly describe cell migration as fluid flow in a viscoelastic solid.

The stability is determined by the sign of the osmotic diffusion constant. It is given in the isotropic case by
\ba
\label{eq10}
D_\phi&=D_2\left(1+l_\phi q^2\right)+\gamma\frac{1-\phi_0}{\phi_0}\left(\frac{4}{3}G+B\right)-\gamma\phi_0\bar{\zeta}=D_1 + D_2 \left(l_\phi q\right)^2.
\end{eqnarray}
Here $D_2=\gamma\phi_0\left(1-\phi_0\right)\chi^{-1}$ is the diffusion constant in the absence of elasticity and activity, given in terms of the inverse osmotic susceptibility $\chi^{-1}=\partial\bar{\mu}/\partial\phi$. $l_\phi=\sqrt{2\kappa\chi}$ is the correlation length due to the interfacial tension, i.e., the width of interfaces in the simple binary-mixture case. The $D_2 l_\phi^2 q^2$ term ensures stability for large wave vectors.

 The second term accounts for elasticity that drives diffusion in order to relax stresses and network strain~\cite{Tanaka1979}. The last term in Eq.~(\ref{eq11}) results from the active solvent stress and can make the diffusion coefficient negative. This is the case for a contractile stress, $\bar{\zeta}>0$. The network is then further contracted in cell-rich regions, where it should extend. The $\bar{\zeta}$ term is equivalent to an active relative force proportional to $\partial_\alpha \phi$ (see~\ref{AppB}), which shifts the osmotic susceptibility and can result in a negative diffusion constant. This is the mechanism described by Murray, Oster, and Harris~\cite{Murray1983,Oster1983}. Our theory in the isotropic case differs from their work mainly because of the global incompressibility that relates the osmotic diffusion constant to elasticity.

Note that the cooperative osmotic diffusion constant is different from the cell self-diffusion constant.  The latter describes correlations in the single-cell velocity, while the former describes correlations in the relative current that depends also on concentration. Alternatively, these diffusion constants are different because the random motion of cells does not necessarily result in concentration changes.

The question is whether the osmotic diffusion constant can become negative for reasonable values of the physical parameters of the cells in the ECM. We examine the different contributions to $D_1$ for $q=0$. They are all proportional to the mobility, multiplied by different energy-density scales: $\chi^{-1}$,\,$G$ and $B$,\,and the active stress $\bar{\zeta}$. We estimate (see~\ref{AppD}) $\chi^{-1}$ and the active stresses to be of order $0.1\,$kPa. The elastic moduli of the ECM, on the other hand, can range between $0.1$ and $10$\,kPa~\cite{Levental2007,Ray2018}. This means that $D_1<0$ is possible only for flexible networks with moduli of the order of $0.1$\,kPa. Note that in the polar case, even for $D_1<0$, other mechanisms can stabilize the system (see Sec.~\ref{sec7}).

\section{Rigid case: matrix stiffness always stabilizes the ECM, while strain-polarization alignment may destabilize it.}
\label{sec6}
The osmotic diffusion coefficient depends on the elastic moduli. For large moduli, it scales as $\sim\gamma G$ and suppresses concentration gradients. This infers stability in the isotropic case, but not necessarily in the polar case. We verify whether alignment mechanisms can destabilize the system in this limit or not.

In the  rigid limit, one solution to the dispersion relation is simply $s=-D_\phi/l_\eta^2$ (see ~\ref{AppC}). This corresponds to stable concentration fluctuations with a decay rate that is comparable with the largest of $\tau$ and $\bar{\tau}$. The other solutions solve
\ba
\label{eq11}
0&=\left(\tau s\right)^2+\left[1+\tau\left(\lambda j_0+D_pq^2-\frac{1}{2}\phi_0^2\left(1-\phi_0\right)^2\frac{\psi}{G}\bar{\psi}\right)\right]\tau s\nonumber\\
&+\tau\left(\lambda j_0+D_pq^2+\bar{\psi}\,u_p (0)\right)
\end{eqnarray}
The system is unstable if either the constant term or the linear coefficient of the quadratic equation is negative.

We examine the signs of the different contributions in Eq.~(\ref{eq11}). The $\lambda j_0$ term is expected to be positive. The steady-state current $j_0<0$ for cells that move in the direction of their polarization, and $\lambda<0$ for cells that align with their direction of motion. The angular diffusion term $D_p q^2$ is also positive. As we consider passive alignment with $\bar{\psi}=-2\psi/\gamma_1$,  the last term is positive as well.  All together, this yields a positive linear coefficient.

The remaining term is $\bar{\psi} u_p(0)$ in the constant term, where $u_p(0)$ describes network displacement in the $x$-direction due to polarization changes. It is purely active and given by
\ba
\label{eq12}
u_p(0)=\frac{1}{2}\phi_0\left(1-\phi_0\right)\frac{\left(1-\phi_0\right)\zeta+\gamma_1\lambda j_0}{G}.
\end{eqnarray}
 These terms are active components of the shear stress. The first stems from the active solvent nematic stress and the latter from the convective polarization stress (shear alignment) in the presence of permeation-alignment and an active current.
 In the rigid limit of large $G$, $u_p(0)$ is negligible. This means that the constant term is positive as well, and the system is stable.

 The strain-polarization coupling can still have an effect in the rigid limit, as long as
 $\psi/G$ is of finite magnitude. This is possible. For nematic elastomers, for example, $\psi/G$ can be related to a typical angle between segments, while $G$ is related to the number of crosslinks~\cite{Warner2007liquid}. As this ratio is at most of order unity within linear elasticity, we use hereafter the value $\phi_0\left(1-\phi\right)\psi\approx-0.1G$.  According to Eq.~(\ref{eq12}),  $u_p(0)>0$, unless the cells are sufficiently extensile ($\zeta<0$). While active cellular stresses are partially contractile ($\zeta>0$) due to the  stresses in the cytoskeleton, they can still be overall extensile, as a result of anisotropic cell division~\cite{Ranft2010} and intercellular interactions~\cite{Balasubramaniam2021}. For reasonable values of the active stress and migration velocity (see~\ref{AppD}), $u_p(0)$ can become negative only for sufficiently small values of the permeation-alignment coupling $|\lambda|<2\times10^{-2}/\mu$m. Furthermore, in order for the constant term in Eq.~(\ref{eq11}) to become negative, the permeation alignment should be even smaller $|\lambda|<2\times10^{-3}/\mu$m. In this case, the instability occurs for small wave vectors up until it is stabilized by angular diffusion. 

  The ECM, therefore, is expected to be stable in the rigid limit, unless three conditions are fulfilled:  a) cells are sufficiently extensile; b) the strain-polarization coupling is of the same order of magnitude as the matrix stiffness; c) the permeation-alignment parameter is small in absolute value.  Note that the third alignment mechanism, concentration alignment, is negligible in this limit, because concentration gradients are suppressed by the large osmotic diffusion coefficient.

 We estimate the relaxation time in the stable case. The retardation time for collagen gels that mimic the ECM are of order of minutes~\cite{Clark2020}. The polarization rates that appear in Eq.~(\ref{eq11}) are of order of h$^{-1}$ (see ~\ref{AppD}). This allows to expand Eq.~(\ref{eq11}) to find $s\approx-1/\tau$ and $s\approx-\left(\lambda j_0+D_pq^2+\bar{\psi}\,u_p (0)\right)$. Together with the pure concentration mode, this means that two modes decay on the scale of minutes, and a third mode that is related to the polarization on the scale of hours.

\section{Analysis in the general case}
\label{sec7}
We analyze the stability in the general case and focus on the stability-instability crossover with zero frequency, $s=0$. The retardation times, which enter the theory via terms linear in $s$, do not play any role in this analysis. The determinant of the linear system in Eqs.~(\ref{eq8}) and (\ref{eq9}) is then given by a quadratic equation, $s^2+2B(q)s+C(q)=0$. The $s=0$ crossover is defined by setting $C=0$. Explicitly, this condition is given by
\ba
\label{eq13}
\frac{l_\phi^4 k_\phi}{D_2 D_p} \left(\bar{\psi}-\lambda j_u\right)u_p(0)+ax+bx^2+x^3&=0,
\end{eqnarray}
where $x=l_\phi^2q^2$. The system is unstable when the left-hand side is negative. The coefficients are given by
\ba
\label{eq14}
\fl a&=\frac{l_\phi^2k_\phi}{D_2}\left[1+\lambda\left(\lambda-l_p^{-1}\right)l_{\gamma1}^2+\frac{D_1}{D_p}\frac{\bar{\psi} u_p(0)+\lambda j_0}{k_\phi}-\frac{\left(1+\lambda^2l_{\gamma1}^2\right)j_0+\left(j_u+\lambda l_{\gamma1}^2\bar{\psi} \right)u_p(0)}{l_p k_\phi}\right],\nonumber\\
\fl b&=\frac{l_\phi^2}{D_p}\left(\bar{\psi} u_p(0)+\lambda j_0\right)+\frac{D_1-l_{\gamma1}^2l_p^{-2}D_p}{D_2}.
\end{eqnarray}
Here we have defined $D_\phi=D_1+D_2 l_\phi^2q^2-\lambda l_p^{-1}l_{\gamma 1}^2D_p.$ $D_1$ is given by the $q^0$ terms in Eq.~(\ref{eq10}),  with slight modifications due to the strain-polarization coupling (see~\ref{AppB}). Note that the $x^3$ interfacial-tension term stabilizes the system for large wave vectors.

\subsection{Strain-polarization coupling stabilizes (destabilizes) contractile (extensile) cells in the ECM for small wave vectors.}
\label{ssec71}
The stability for small wave vectors is determined by the sign of the constant term in Eq.~(\ref{eq13}) $\sim k_\phi\left(\bar{\psi}-\lambda j_u\right)u_p(0)$. This term is active. It occurs because both the concentration and polarization are not pure hydrodynamic modes and have a finite relaxation rate for $q=0$. The concentration, which is a conserved quantity in passive systems, has a finite relaxation rate due to cell division. The polarization, which is usually a soft mode, is coupled to the strain and has a finite relaxation/growth rate due to active shear stresses that polarization rotation imposes on the network.

The question is whether this constant term stabilizes the ECM or destabilizes it. As was shown above, cells that align with matrix segments yield $\bar{\psi}>0$. Contractile cells have $u_p(0)>0$, while sufficiently extensile cells have $u_p(0)<0$. The remaining term to examine is the strain-induced relative current, $j_u=\gamma\left(2\lambda\phi_0\left(1-\phi_0\right)\psi-\nu'\right)$. The first term is positive for parallel cell-network alignment ($\psi<0$) and cells that align with their direction of motion ($\lambda<0$). The $\nu'$ term is expected to be negative, similarly to $\nu$, such that the cells flow in their direction of polarization. This can also be attributed to a strain-dependent friction coefficient (see~\ref{AppB}). This yields $j_u>0$.

\begin{figure}[ht]
\centering
\includegraphics[width=0.8\textwidth]{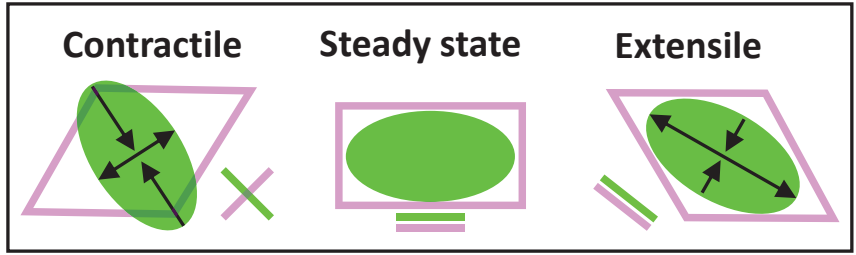}
\caption{(Color online) Heuristic description of matrix mesh (purple) deformation by the rotation of active, nematic cells (green). The black arrows indicate the direction of forces exerted by the cells. At steady state (center), the cells and matrix are aligned. As contractile cells rotate (left), the matrix deforms and aligns normal to the cells. As extensile cells rotate (right), the matrix deforms and aligns parallel to the cells. Cell and matrix orientations are indicated by the small green and purple lines.}
\label{fig1b}
\end{figure}

Overall, the constant term has the same sign as $u_p(0)$.  This means that the net effect of cell division and several alignment mechanisms depends mostly on the sign of the active stress. For contractile cells and weakly extensile cells, the combined effects facilitate a homogeneous, steady flow. For sufficiently extensile cells, an instability can occur, during which large domains of different concentrations and polarizations form. A similar version of this instability was described in Ref.~\cite{Maitra2018} for active, uniaxial, elastomeric gels. The mechanism behind this instability is nematic in nature. It can be understood heuristically by considering  a fluctuation in the orientation of active nematic cells (Fig.~\ref{fig1b}).  As the cells rotate, the mesh deforms elastically in order to balance the cellular active stress. Extensile cells deform the mesh in a way that aligns it parallel with the cells. This drives the rotation of additional cells, because of the aligning interaction. Such a positive feedback infers an instability. Contractile cells, on the other hand, deform the mesh in a way that aligns it perpendicularly to the cells. This drives alignment in the normal direction and relaxes the fluctuation.

The scaling of the domain size in the unstable case depends on other system parameters. In the simple case where $a>0$, the most unstable mode is $q=0$ and system-size domains form for sufficiently large systems. For a negligible permeation-alignment coupling ($\lambda=0$), the critical system size is $L_c=l_a/\sqrt{-\phi_0\left(1-\phi_0\right)\psi/G},$ where $l_a=2\pi\sqrt{K/|\zeta|}$ is the active length.  A similar scaling appears in the more general analysis below (see~Sec.~\ref{ssec74}), as well as other active flow instabilities in nematic cells~\cite{Duclos2018}. Reasonable values of the physical parameters (see~\ref{AppD}) yield $L_c$ of the order of $100\,\mu$m. The growth rate is found from the $q=0$ limit of Eq.~(\ref{eq9}). It can be approximated as $\phi_0\left(1-\phi_0\right)\psi\zeta/\left(G\gamma_1\right).$ Our estimates (see~\ref{AppD}) yield a growth time of approximately $100$\,h. It is shorter for extensile active stresses that are larger than $0.1$kPa or more negative values of $\phi\left(1-\phi\right)\psi/G<-0.1$. The latter limit infers strains of order $1$ and a quantitative treatment of it would require a framework of nonlinear elasticity.

Extensile cells can induce an instability also via the term $\sim \bar{\psi}u_p(0) D_1/D_p$ in $a$. This requires large $D_1/D_p$ values and was described as part of the rigid-limit analysis in Sec.~\ref{sec6}. Below, we focus on the contractile case.

\subsection{Concentration alignment can stabilize or destabilize the ECM.}
\label{ssec72}
We now analyze Eq.~(\ref{eq13}) for arbitrary $q$ values. The system is unstable when LHS is negative. As $x>0$, this requires either $a<0$ or $b<0$. Reviewing Eq.~(\ref{eq14}), we find that the only possible source of instability, aside from $D_1<0$, is concentration alignment.

Concentration alignment can destabilize the system via three mechanisms. The first  mechanism is the term $\lambda\left(\lambda-l_{p}^{-1}\right)$ in $a$ that originates from the effective angular diffusion coefficient, $\left[1+\lambda\left(\lambda-l_{p}^{-1}\right)l_{\gamma1}^2\right]D_{p}$. For sufficiently negative $l_p^{-1}$, the angular diffusion coefficient becomes negative. While this is a passive mechanism, it has an effect only in the active case where $k_\phi>0$.

The second mechanism is the term $\sim l_{p}^{-1}\left(\lambda-l_{p}^{-1}\right)$ in $b$. This is a known passive instability mechanism, where the concentration-alignment coupling favors gradients in concentration and polarization~\cite{Blankschtein1985,Hinshaw1988,Voituriez2006} over a homogeneous state. It depends on $l_p^{-2}$ and not on the sign of $l_p$.

The third mechanism is described by the last terms in $a$ in Eq.~(\ref{eq14}). The $j_0$ term is responsible for the active instability reported in Ref.~\cite{Adar2021}. Consider a concentration fluctuation. The sign of $l_p$ determines whether cells orient into or out of cell-rich regions ($l_{p}<0$ or $l_{p}>0$, respectively, and see Fig.~\ref{fig1}). Cells with $l_p<0$ actively flow in the direction of their polarization, resulting in an instability.

The $l_p>0$ case may also become unstable due to the strain-induced current, $j_u>0$. For sufficiently large $j_u$, cells would move in the $yz$ plane oppositely to how they re-orient. However, 
for reasonable physical values (see ~\ref{AppD}), the $j_u$ term is negligible compared with the $j_0$ term. Overall, $l_p<0$ is  expected to be destabilizing, while $l_p>0$ is expected to be stabilizing, except for when $D_p l_{\gamma1}^2l_p^{-2}>D_1$.

\subsection{Stability diagrams}
\label{ssec73}
So far we have identified destabilizing terms. We now precise the instability condition. Eq.~(\ref{eq13}) is a cubic equation with positive free and cubic coefficients.  For it to become negative, it must have a minimum point for some $x_0=\left(q_0 l_\phi\right)^2>0$ that has a negative value, as is illustrated in Fig.~\ref{fig3}(a). Equating $C'(q_0)=0$ yields $x_0=-b/3\pm\sqrt{\left(b/3\right)^2-a/3}$. We further require its value to be negative, i.e., $C(q_0)<0$.

These conditions enable to determine the stability of the system. We focus on flexible matrices, such that an instability is possible, and substitute reasonable physical values for cells in the ECM. An important variable to take into account is the mobility $\gamma$ that appears in most of the terms in Eqs.~(\ref{eq13}) and (\ref{eq14}). It is related to the matrix architecture and is expected to scale as $\gamma\sim\xi^2$. Namely, the active alignment mechanism $\sim j_0/l_p$ in the $a$ term of Eq.~(\ref{eq14}) is more important for smaller $\xi$ values. As $\xi$ becomes larger, its contributions becomes negligible with respect to osmotic diffusion and permeation alignment.

\begin{figure*}[ht]
\includegraphics[width=0.32\textwidth]{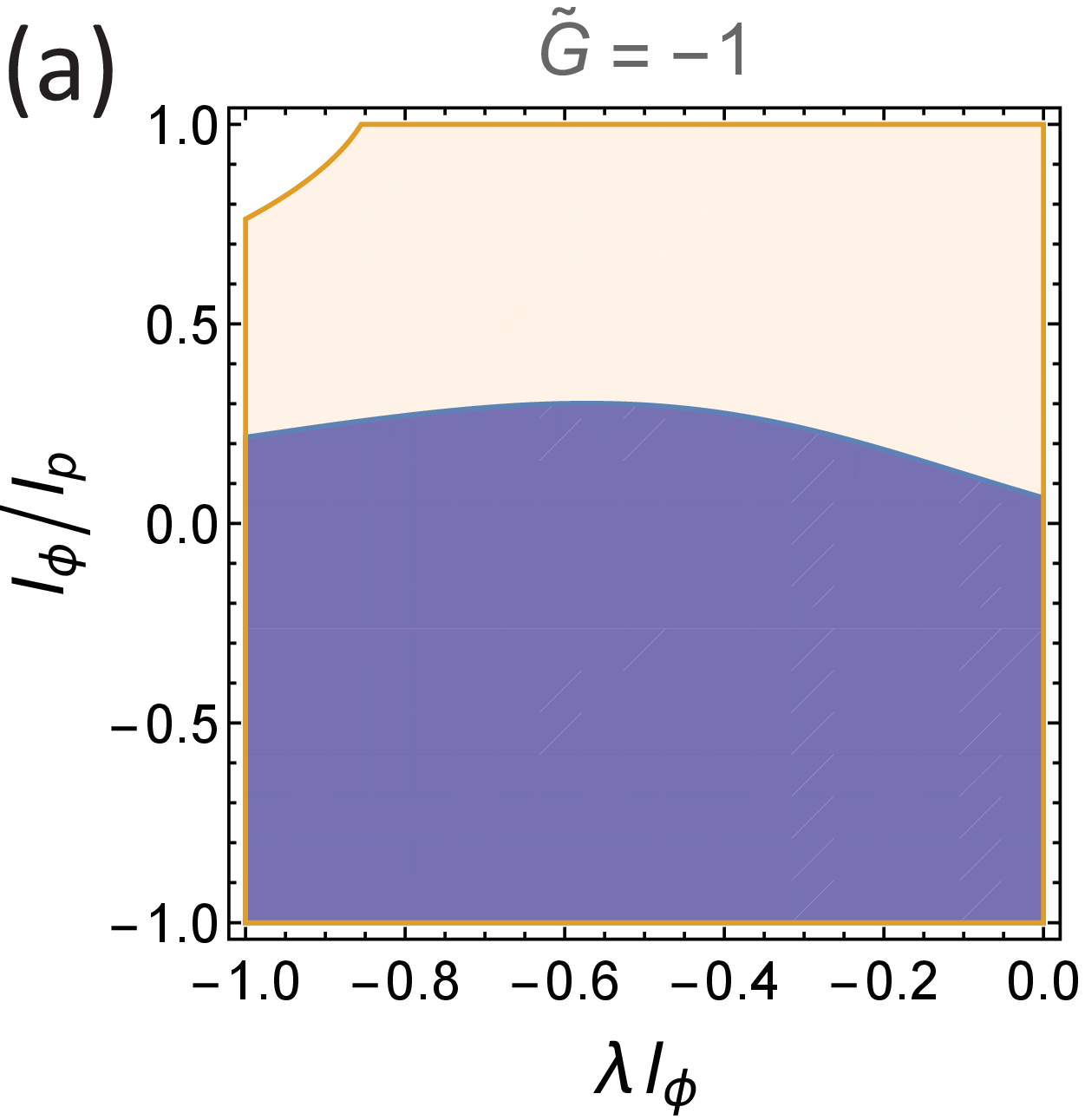}
\includegraphics[width=0.32\textwidth]{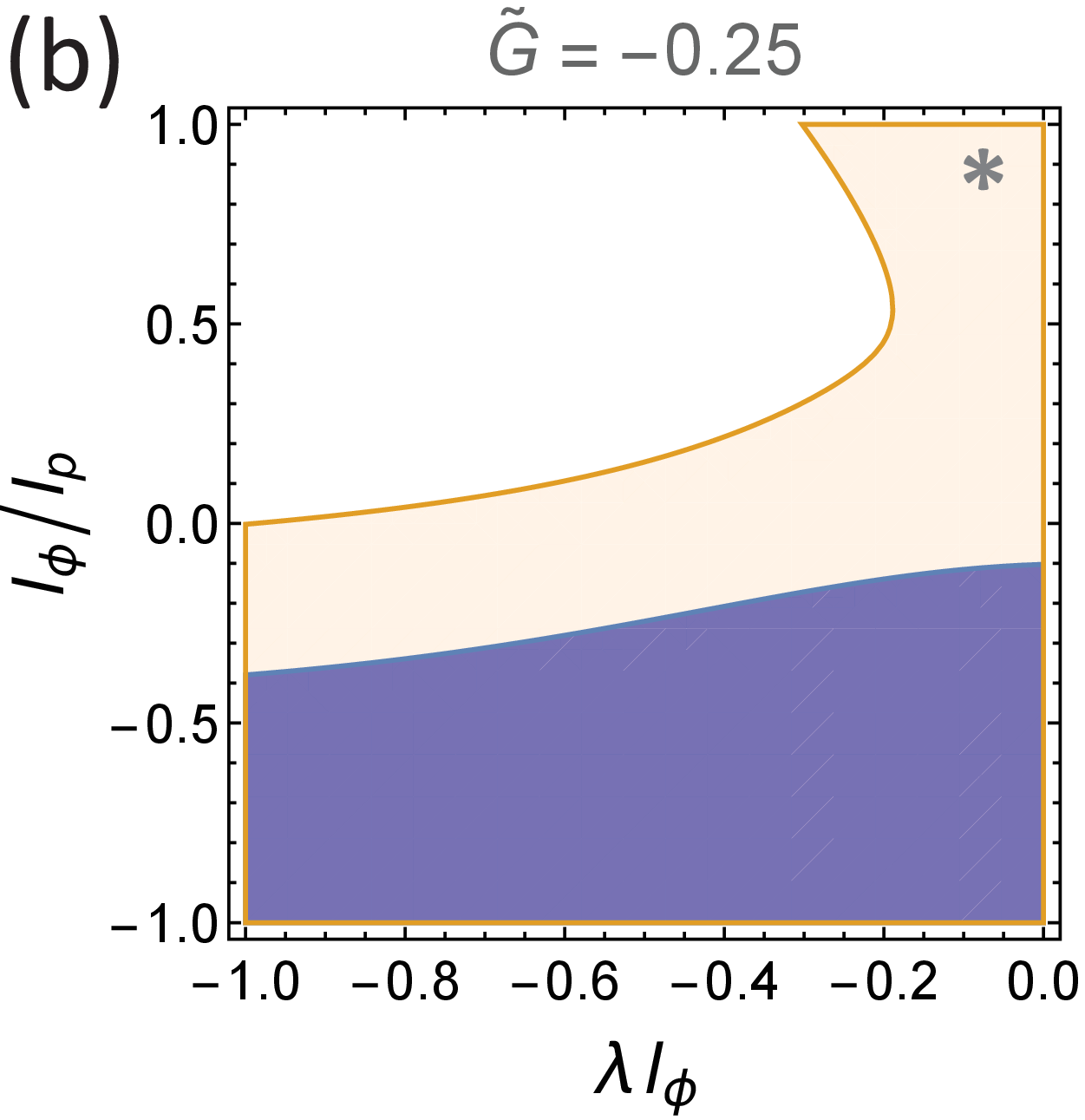}
\includegraphics[width=0.32\textwidth]{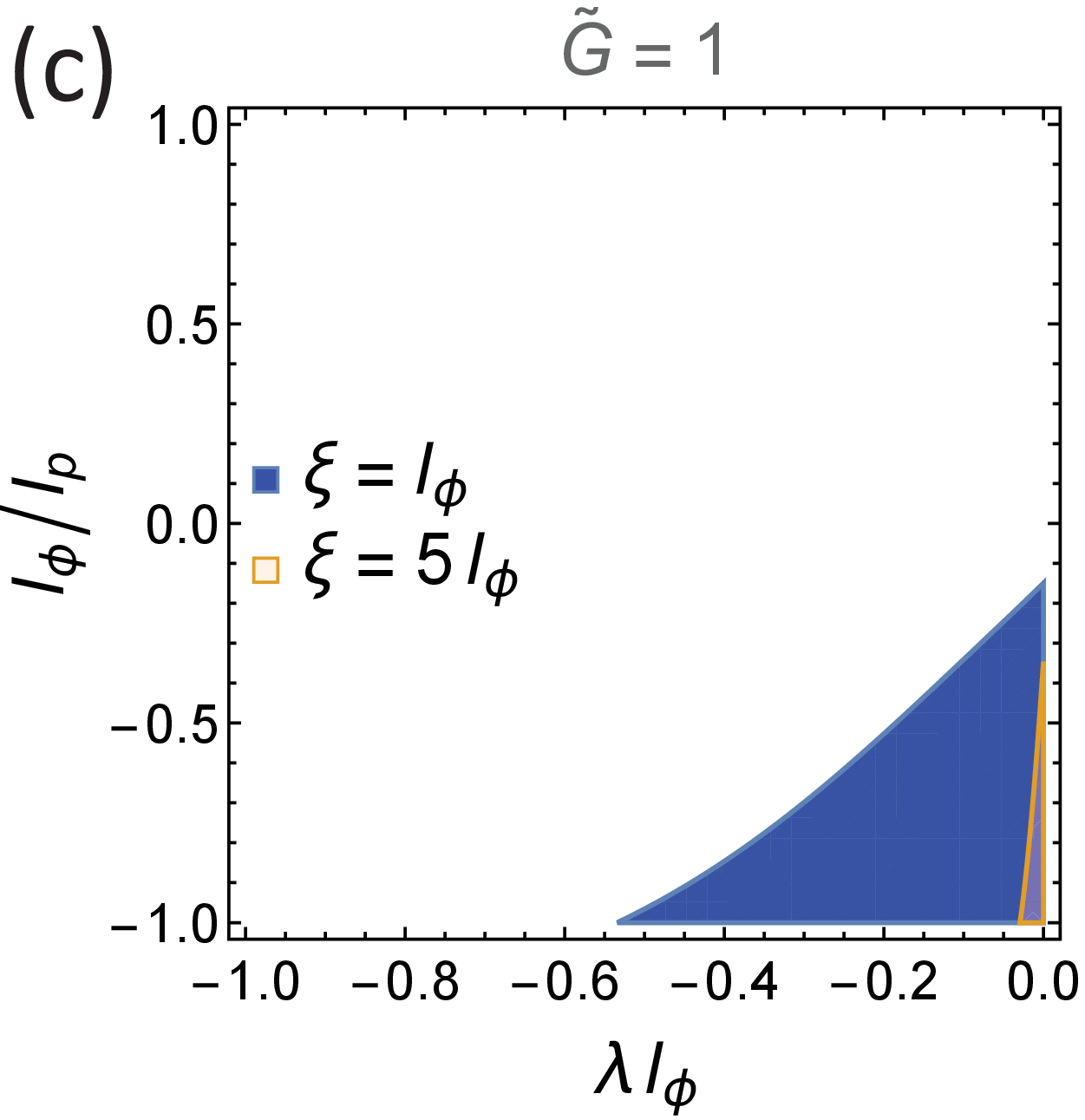}
\caption{(Color online). Stability diagrams for $\xi=l_\phi$ (blue) and $\xi=5l_\phi$ (light orange) and different osmotic diffusion constants $\sim\gamma\tilde{G}$. Colored regions are unstable. The values $D_p=2.5\,l_\phi^2 k_\phi$, $\bar{\psi}=0.25\,k_\phi$, $j_0=-10l_\phi k_\phi$, and $j_u=\left(10 l_\phi-0.25\lambda\xi^2\right)k_\phi$ are used, in accordance with the estimates of~\ref{AppD}. The asterisk sign in Fig. (b) marks the region where the passive concentration-alignment instability takes place. The other regions correspond to active instabilities.}
\label{fig2}
\end{figure*}

Our results are presented in Fig.~\ref{fig2} using stability diagrams in the $\left(\lambda l_\phi,\,l_\phi/l_p\right)$ parameter space. We define $D_1=\gamma\tilde{G}K/l_\phi^2$ and draw three different diagrams for three values of the dimensionless $\tilde{G}$.
Each diagram illustrates stable (white) and unstable (colored) regions for two different mesh sizes, $\xi/l_\phi=1,\,5$.

Figures~\ref{fig2}(a) and (b) show that the system is generally unstable for a negative osmotic diffusion constant. The system may still be stable for sufficiently large $l_p^{-1}$ and negative $\lambda$ values. This is thanks to the stabilizing couplings to the active relative current and the large effective angular diffusion coefficient. The system is harder to stabilize for more negative $\tilde{G}$ values and larger $\xi$ values, which yield a more negative osmotic diffusion constant. The region in the top right corner of Fig.~\ref{fig2} (b), which is marked with an asterisk sign, is unstable due to the passive concentration-alignment mechanism that occurs for $D_p l_{\gamma1}^2l_p^{-2}>D_1$. This region remains unstable for  $\tilde{G}<0.5$. Figure~\ref{fig2}(c) demonstrates that the system is relatively stable for $\tilde{G}>0$. The system is more susceptible for instabilities for $\xi=l_\phi$, where the stabilizing osmotic diffusion coefficient is smaller. The instability is the active concentration-alignment instability of Sec.~\ref{ssec72}. Note that it occurs for small $\lambda$ values in absolute values. For more negative $\lambda$ values, the permeation-alignment mechanism stabilizes the system.

\subsection{Critical wave vector}
\label{ssec74}
\begin{figure*}[ht]
\includegraphics[scale=0.39]{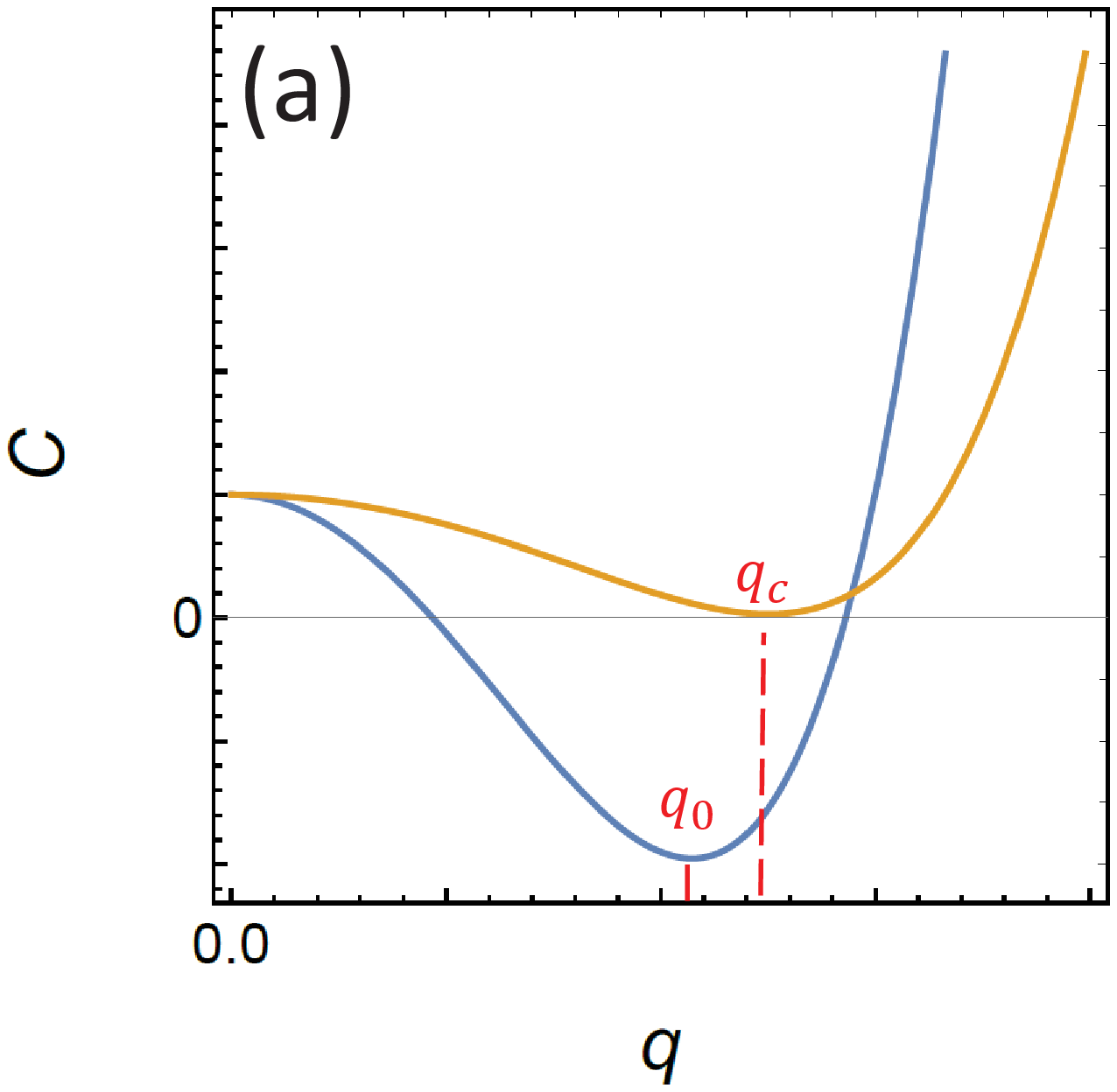}
\includegraphics[scale=0.5]{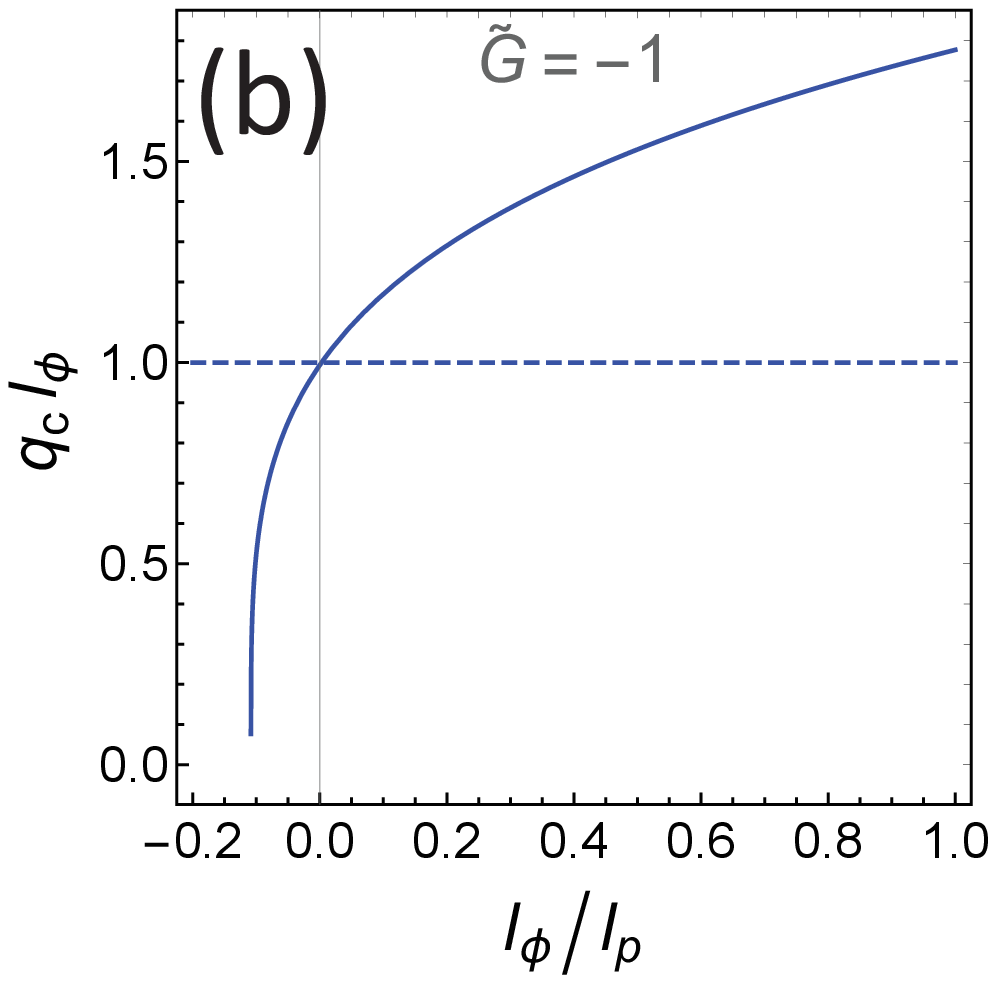}
\includegraphics[scale=0.53]{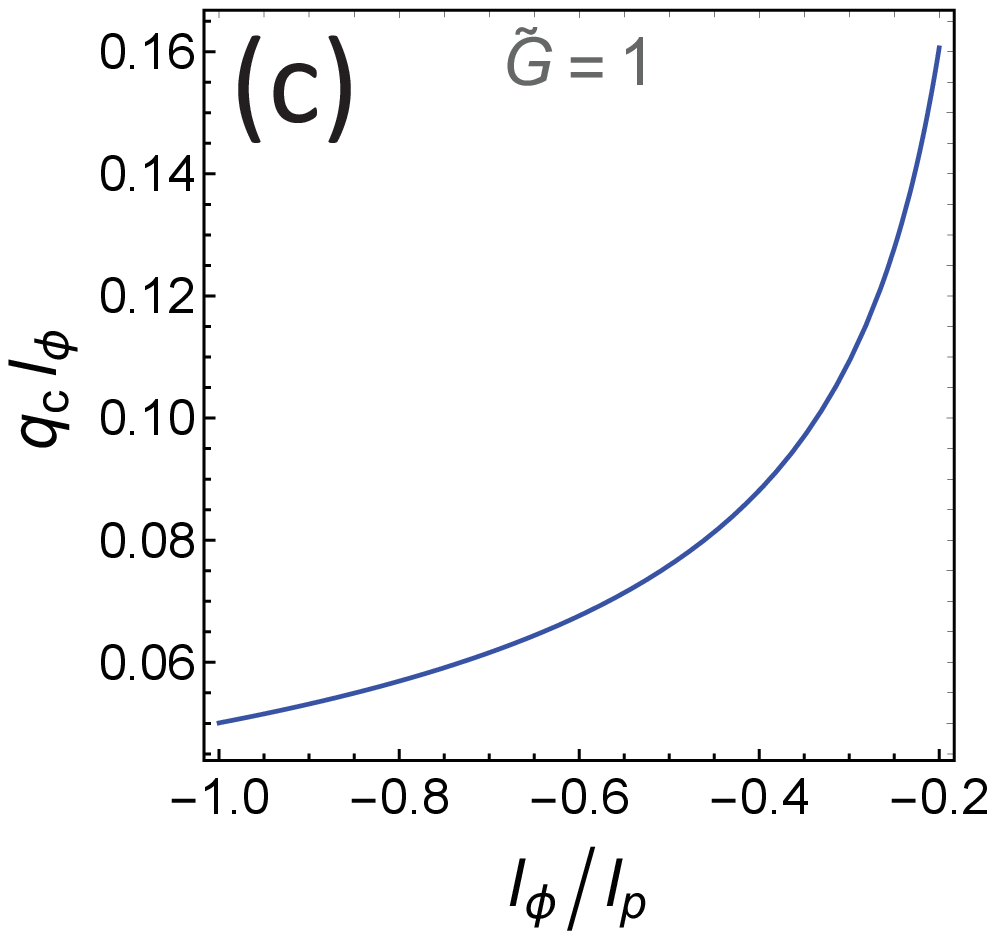}
\caption{(Color online). (a) $C(q)$ in arbitrary units. The system is unstable for $C<0$. The wave vector $q_0$ is found as a negative, minimal point, while the critical wave vector $q_c$ is a degenerate root. (b) Critical wave vector as a function of $l_p^{-1}$ in the case $a>0$ for $\tilde{G}=-1$. The result in the isotropic case ($l_p^{-1}=0$) is marked by a dashed line. (c) Critical wave vector as a function of $l_p^{-1}$ in the case $a<0$ for $\tilde{G}=1$. The same values as those of Fig.~\ref{fig2} are used with $\xi=l_\phi$. }
\label{fig3}
\end{figure*}

As the system is stable for both vanishing and large wave vectors, all the aforementioned instabilities occur at finite wave vectors. At the critical system parameters, there is only one marginally stable, critical wave vector $q_c$. This is illustrated in Fig.~\ref{fig3}(a). The critical wave vector is found from $C(q_c)=C'(q_c)=0$, where $C(q)$ is the polynomial in Eq.~(\ref{eq13}).

We find $q_c$ in the reasonable limit where the constant term of Eq.~(\ref{eq13}) is small, i.e., $0<\frac{l_\phi^4}{D_2 D_p}k_\phi \left(\bar{\psi}-\lambda j_u\right)u_p(0)\ll1$. The solution depends on the sign of $a$ that is defined in Eq.~(\ref{eq14}). It is given by
\ba
\label{eq15}
q_c l_\phi=a^{1/4}&\qquad a>0,\nonumber\\
q_c l_\phi=\left(-2\frac{l_\phi^4 k_\phi}{D_2 D_p} \left(\bar{\psi}-\lambda j_u\right)\frac{u_p(0)}{a}\right)^{1/2}&\qquad a<0.
\end{eqnarray}
In the isotropic limit, $a=l_{\phi}^{2}k_{\phi}/D_{2}$, and our result reduces to that of Oster et al.~\cite{Murray1983,Oster1983}. This is a generic scaling for phase separation of reproducing entities (see also Ref.~\cite{Cates2010} for pattern formation in bacteria). In the polar case, while the scaling $q\sim\left(k_{\phi}/D_{2}\right)^{1/4}$ still holds, the prefactor can change substantially due to alignment mechanisms. This is illustrated in Fig.~\ref{fig3}(b) as a function of $l_p$ for $\lambda=0$.  $a$ decreases as $l_p$ becomes more negative and, consequently, the critical wave vector decreases as well. It becomes infinitesimally small as $a$ approaches zero.

The second line in Eq.~(\ref{eq15}) arises due to the active concentration-alignment mechanism. We focus on the $\lambda=0$ case and plot the critical wave vector for different $l_p$ values in Fig.~\ref{fig3}(c). It is possible to estimate it in the limit where $-j_{0}/\left(l_{p}k_{\phi}\right)$ is the dominant contribution to $a$. This yields the scaling
\be
\label{eq16}
q_c\sim\sqrt{-\frac{\psi}{G}\frac{l_p k_\phi}{j_0}\frac{\zeta}{K}}.
\ee
Namely, the critical wavelength is proportional to the active length $l_a$. The critical wave vector is expected to be small in this limit, as is evident from Fig.~\ref{fig3}(c).

In addition to the critical wave vector, there is also a fastest growing mode with wave vector $q_\ast$, for which $s$ is maximal. This infers the formation of transient periodic domains of size $\sim 1/q_\ast$ with continuous flow patterns between the domains. This may  correspond to the formation of clusters or strands of cells that migrate in localized collections. In standard binary systems, these domains grow into macroscopically phase-separated regions. However, any coarsening dynamics is arrested by cell division and death~\cite{Cates2010}, which does not allow stable macroscopic domains with a concentration that is different from the homeostatic one $\phi\ne\phi_0$.
%

\section{Discussion}
\label{sec8}
In this paper, we have formulated an active-gel theory to describe multicellular migration in the ECM as an active, polar solvent permeating in a viscoelastic solid. The theory accounts naturally for dynamic reciprocity and classifies clearly different cell-matrix interactions. Namely, we highlight three alignment mechanisms that relate the polarization with the strain ($\sim\psi$), permeation current ($\sim\lambda$) and concentration gradients ($\sim l_p^{-1}$). The three are distinguishable; The $\psi$ coupling, unlike the other two, occurs also for nematic cells and in the absence of concentration gradients and current. It is possible to separate between the permeation- and concentration-alignment mechanisms by studying cells with different motilities and in different setups (e.g., in the bulk of a cell collection, compared to an invading cell front).

A main conclusion of our work regards the effect of cell-matrix alignment on the stability of homogeneous multicellular migration for small wave vectors. Confluent cell monolayers can be extensile or contractile depending on the cell type and the regulation of cell-cell adhesions~\cite{Balasubramaniam2021}. Our results indicate that cells with different values of the active stress would migrate in qualitatively different manners within a three-dimensional matrix. Contractile and weakly extensile cells flow homogeneously, while sufficiently extensile cells form domains. This simple distinction is remarkable, given the large number of forces and alignment mechanisms that take place. We note that cell-cell interactions in polarized cells that migrate in a fluid-like manner are generally weaker than those in confluent monolayers.

	
In addition to this strain-driven instability in the extensile case, our analysis suggests two possible origins for instability in the contractile case: a negative, effective diffusion constant, due to active forces, or a negative $l_p<0$ that aligns cells towards larger cellular concentrations. The value of $l_p$ tunes the critical wave vector of the instability.

While this work focuses on the linear stability of homogeneous polarized cells, our framework could be equally useful in other relevant situations. It could describe, for example, the flow of cell fronts during invasion or an isotropic-polar transition of cells in the ECM. The combination of alignment mechanisms is expected to polarize cells. Namely, cells would flow mainly parallel to network segments, due to anisotropy of friction coefficient, while the polarization and migration speed are expected to feedback via the permeation-alignment mechanism, resulting in polarized, flowing cells.

Another interesting example is cells migrating in tracks of aligned collagen fibers~\cite{Clark2015,Han2016}. Concentration gradients are expected to be small in this case, and the flow in the normal direction to the polarization is expected to be negligible. Most of the effects discussed in this work should not be relevant then. This is evident by taking the limit of vanishing mobility. Rather, the cells could be described in this case as a fluid flowing in a confined geometry, similarly to the theoretical description of in-vitro migration experiments in channels (see, e.g., ~\cite{Duclos2018}). As cell-matrix friction becomes a boundary effect, the cellular viscous stresses become important in this setup.

Several extensions of our theory can be considered. It is possible to add components to the current two-component description that coarse-grains the solvent and cells together. This overlooks solvent-cell friction, which is reasonable, because the cells are almost ten orders of magnitude more viscous that water and thus dominate the dissipative processes. The description becomes problematic, however, when the cellular concentration is inhomogeneous across the solvent. For example, when contractile cells are added to a pre-existing gel, they contract it and squeeze out some of the solvent. This can be described in our two-component theory only in an indirect way, by varying the value of the active stresses. Second, this framework overlooks diverse cell species, such as cancer cells vs. fibroblasts. Fibroblasts are especially interesting because they are abundant in the ECM and are able to remodel it. Active matrix remodeling is currently missing from our theory. The theory can also be adapted for more complicated rheological descriptions of the ECM~\cite{Elosegui-Artola2021}, including plasticity, long-time stress relaxation and non-linearities. A treatment of the latter two within the same framework can be found in~\cite{Adar2021}, where we considered also reversible network deformation by the migration current(``permeation deformation''). We reserve the further study of  this effect and active remodeling  to a future work.

In conclusion, this work could open an avenue for studying cell-migration modes, ECM patterning, and cell-ECM interactions in three dimensions and at the mesoscopic scale. Thanks to the generic framework,  it can be used to study other physical systems, such as bacteria and active colloids in viscoelastic media. It would be interesting to apply our theory to three-dimensional migration experiments that measure cell concentration, alignment and velocity in the ECM or engineered gels. We suggest to compare between the migration of extensile and contractile cells and to verify whether domains form for sufficiently extensile cells. In addition, measuring and deducing typical values of $l_p$ for different cell types should be important in characterizing their migration modes.

\ack
RMA acknowledges funding from Fondation pour la Recherche Médicale (FRM Postdoctoral Fellowship).

\appendix
\section{Polarization-rate constitutive equation}
\label{AppA}
We make use of the general framework of non-equilibrium thermodynamics,
similarly to Ref.~\cite{Adar2021}.
We start from the general equation for the polarization dynamics
\begin{eqnarray}
\label{eqA1}
\left(\partial_{t}+v_{\beta}^{c}\partial_{\beta}\right)p_{\alpha}-p_{\beta}\partial_{\beta}v_{\alpha}^{c} & =\frac{1}{\gamma_{1}}h_{\alpha}+\lambda j_{\alpha}+\phi\left(1-\phi\right)\psi'\Delta\mu\epsilon_{\alpha\beta}p_{\beta}.\label{eq:seq1}
\end{eqnarray}
The derivative on the left-hand side is a material derivative and
a convective term. The cells are assumed to convect by their own velocity
and rotate due to their strain-rate like rigid rods (shear-alignment
parameter of $-1$). The right-hand side accounts for dissipative
couplings. The coupling to the orientational field $h_{\alpha}=-\delta F/\delta p_{\alpha}$
is given in terms of the rotational viscosity $\gamma_{1}.$ The coupling
to the relative current, $j_{\alpha}$ is referred to as the permeation-alignment
coupling. The coupling to activity $\Delta\mu$ includes the strain
(a term $\sim\Delta\mu p_{\alpha}$ simply renormalizes the parallel
field, $h_{\parallel}$) and is proportional to $\phi\left(1-\phi\right),$
because it involves interaction with the matrix.

The orientational field is given by
\begin{eqnarray}
\label{eqA2}
h_{\alpha} & =h_{\parallel}p_{\alpha}+K\left(\partial_{\beta}\partial_{\beta}p_{\alpha}+l_{p}^{-1}\partial_{\alpha}\phi\right)-2\phi\left(1-\phi\right)\psi\epsilon_{\alpha\beta}p_{\beta}.\label{eq:seq2}
\end{eqnarray}
Substituting in Eq.~(\ref{eqA1})
yields Eq.~(\ref{eq3}) with $\bar{h}_{\parallel}=h_{\parallel}/\gamma_{1}$,
$D_{p}=K/\gamma_{1},$ and $\bar{\psi}=\psi'\Delta\mu-2\psi/\gamma_{1}.$
We consider hereafter $\psi'=0,$ such that strain-polarization alignment is
driven by the passive free-energy coupling, $\bar{\psi}=-2\psi/\gamma_{1}.$

\section{Linearization of the dynamic equations }
\label{AppB}
In this Appendix we derive the linearized version of the equations,
which is used for the linear stability analysis. We first write the
equations in full form, including explicit expressions for the fields
that are derived from the free energy. Then, we solve the steady-state
equations, and linearize around the steady-state solution.

\subsection*{Matrix forces}

The dynamic equations include force-balance equations, written in
terms of the cell and matrix forces. The matrix force is given in
Eq.~(\ref{eq5}). It
includes the divergence of the elastic stress,
\begin{eqnarray}
\label{eqB1}
\sigma_{\alpha\beta}^{el} & =\frac{\delta F}{\delta\epsilon_{\alpha\beta}}=\frac{\phi}{\phi_{0}}\left(2G\tilde{\epsilon}_{\alpha\beta}+B\epsilon\delta_{\alpha\beta}\right)+\phi\left(1-\phi\right)\psi Q_{\alpha\beta}.
\end{eqnarray}
 In addition, it includes the force that results from osmotic pressure
gradients, $-\phi\partial_{\alpha}\overline{\mu}.$ The relative chemical
potential is given by
\begin{eqnarray}
\label{eqB2}
\overline{\mu} & =\frac{\delta F}{\delta\phi}=-k_{B}Ta^{-3}\left(1+\ln\left(1-\phi\right)\right)+\left(1-2\phi\right)\left(\chi_{0}+\psi Q_{\alpha\beta}\epsilon_{\alpha\beta}\right)\nonumber\\&+G\tilde{\epsilon}^{2}+\frac{1}{2}B\epsilon^{2}+Kl_{p}^{-1}\partial_{\alpha}p_{\alpha}-2\kappa\partial_{\beta}\partial_{\beta}\phi.
\end{eqnarray}

\subsection*{Steady state}

We consider a homogeneous steady-state, as is described in Sec.~\ref{sec4}.
For a homogeneous system, the forces acting on the cells and matrix
[Eqs. (\ref{eq5}) and (\ref{eq6})]
vanish. Force balance then requires that the relative force vanishes
as well, i.e.,
\begin{eqnarray}
\label{eqB4}
\frac{1}{\gamma}j_{\alpha}^{0} & =\phi_{0}\left(1-\phi_{0}\right)\left(\lambda h_{\alpha}^{0}+\left[\nu-\nu'\frac{\psi}{3G}\phi_{0}\left(1-\phi_{0}\right)\right]p_{\alpha}^{0}\right).
\end{eqnarray}
The steady-state molecular field results from the active relative and is given by
\begin{eqnarray}
\label{eqB5}
h_{\alpha}^{0} & =-\gamma_{1}\lambda j_{\alpha}^{0}.
\end{eqnarray}
Inserting this
result in the previous equation yields the relative current at steady
state,
\begin{eqnarray}
\label{eqB6}
 j_{\alpha}^{0} & =\frac{\gamma}{1+\lambda^{2}l_{\gamma1}^{2}}\phi_{0}\left(1-\phi_{0}\right)\left[\nu-\nu'\frac{\psi}{3G}\phi_{0}\left(1-\phi_{0}\right)\right]p_{\alpha}^{0},
\end{eqnarray}
where $l_{\gamma1}=\sqrt{\phi_{0}\left(1-\phi_{0}\right)\gamma_{1}\gamma}$
is a screening length due to the interplay between friction and rotational
viscosity. We consider $\nu,\nu'<0,$ such that the cells migrate
in the direction of their polarization. The role of $\nu'$ at
steady state is to renormalize the motility. The permeation-alignment mechanism effectively increases the
friction.

\subsection*{Linearized equations}

The stability is studied by introducing a small perturbation in the
fields at point $r_{\alpha}$ and time $t$ with a wave vector $q_{\alpha}$
and growth rate $s$,
\begin{eqnarray}
\label{eqB7}
\left(\phi,p_{\alpha},u_{\alpha}\right) & \approx\left(\phi_{0},p_{\alpha}^{0},u_{\alpha}^{0}\right)+\left(\phi^{1},p_{\alpha}^{1},u_{\alpha}^{1}\right)\exp\left(iq_{\alpha}r_{\alpha}+st\right).
\end{eqnarray}
For simplicity, we consider $q_{x}=0.$ Also, in order to maintain
the modulus of the polarization, $p_{x}^{1}=0.$ As is explained in
the paper, it is possible to integrate over the displacement variable
and to analyze the stability in terms of $\phi^{1}$ and $p_{d}^{1}=iq_{\alpha}p_{\alpha}^{1}.$

The linearized continuity equation [Eq.~(\ref{eq2})] is given by
\begin{eqnarray}
\label{eqB8}
s\phi^{1} & =-iq_{\alpha}j_{\alpha}^{1}-k_{\phi}\phi^{1}.
\end{eqnarray}
The linearized equation for the polarization [divergence
of Eq.~(\ref{eq3})]
reads
\begin{eqnarray}
\label{eqB9}
sp_{d}^{1} & =\bar{h}_{\parallel}p_{d}^{1}-D_{p}q^{2}p_{d}^{1}-D_{p}l_{p}^{-1}q^{2}\phi^{1}+\lambda iq_{\alpha}j_{\alpha}^{1}\nonumber\\&+\phi_{0}\left(1-\phi_{0}\right)\bar{\psi} iq_{\alpha}\left(\epsilon_{\alpha\beta}p_{\beta}\right)^{1}.
\end{eqnarray}
 The parallel field can be found from the polarization-rate equation
at steady state, $\bar{h}_{\parallel} =-\lambda j_{0}+\frac{1}{3}\phi_{0}^{2}\left(1-\phi_{0}\right)^{2}\bar{\psi}\psi/G$.
 The $iq_{\alpha}\left(\epsilon_{\alpha\beta}p_{\beta}\right)^{1}$
term in Eq.~(\ref{eqB9}) is given by
\begin{eqnarray}
\label{eqB10}
iq_{\alpha}\left(\epsilon_{\alpha\beta}p_{\beta}\right)^{1} & =-\frac{1}{2}q^{2}u_{x}^{1}+\frac{\psi}{6G}\phi_{0}\left(1-\phi_{0}\right)p_{d}^{1}.
\end{eqnarray}
The displacement in the $x$-direction is found from the force balance
on the entire gel [sum of the two lines in Eq.~(\ref{eq4})] in the $x$-direction. As $q_{x}=0,$ only the total
shear stress of the system contributes to this force. We find that
\begin{eqnarray}
\label{eqB11}
u_{x}^{1} & =\frac{1}{G\left(1+\tau s\right)q^{2}}\left[\left(1-\phi_{0}\right)\zeta+\phi_{0}\left(1-\phi_{0}\right)\psi-h_{x}^{0}\right]p_{d}^{1}.
\end{eqnarray}
This demonstrates how the network is strained by active stresses ($\zeta$
term) and alignment mechanisms ($h_{x}^{0}$ term), as well as the stress due to passive alignment
($\psi$ term).

We find that
\begin{eqnarray}
\label{eqB12}
iq_{\alpha}\left(\epsilon_{\alpha\beta}p_{\beta}\right)^{1} & =-\frac{1}{\phi_{0}\left(1-\phi_{0}\right)}\left(u_{p}+\frac{1}{3}\phi_{0}^{2}\left(1-\phi\right)^{2}\frac{\psi}{G}\right)p_{d}^{1},
\end{eqnarray}
 where $u_{p}$ is given by
\begin{eqnarray}
\label{eqB13}
u_{p}&=u_{p}\left(0\right)-\frac{\tau s}{1+\tau s}\left(u_{p}\left(0\right)+\frac{1}{2}\phi_{0}^{2}\left(1-\phi_{0}\right)^{2}\frac{\psi}{G}\right),\nonumber \\
u_{p}\left(0\right)& =\frac{\phi_{0}\left(1-\phi_{0}\right)}{2G}\left[\left(1-\phi_{0}\right)\zeta-h_{x}^{0}\right].
\end{eqnarray}
 This yields overall
\begin{eqnarray}
\label{eqB14}
sp_{d}^{1} & =-\left(\lambda j_{0}+\bar{\psi} u_{p}+D_{p}q^{2}\right)p_{d}^{1}-D_{p}l_{p}^{-1}q^{2}\phi^{1}+\lambda iq_{\alpha}j_{\alpha}^{1}.
\end{eqnarray}

It remains to find the divergence of the relative current. We take
a linear combination of the two force balance equations in Eq.~(\ref{eq4})
of the paper, and find that
\begin{eqnarray}
\label{eqB15}
\fl \frac{1}{\gamma}iq_{\alpha}j_{\alpha}^{1} & =iq_{\alpha}\left(\phi_{0}\left(1-\phi_{0}\right)\left[\lambda h_{\alpha}^{1}+\nu p_{\alpha}^{1}+\nu'\left(\epsilon_{\alpha\beta}p_{\beta}\right)^{1}\right]+\left(1-\phi_{0}\right)f_{\alpha}^{m1}-\phi_{0}f_{\alpha}^{c1}\right).
\end{eqnarray}
Before resuming the calculation, we note that the active relative
force $\sim\nu'$ plays, in part, a similar role to an anisotropic
friction coefficient. To see this, consider a friction coefficient
$\gamma_{\alpha\beta}^{-1}=\gamma_{0}^{-1}\delta_{\alpha\beta}+\gamma_{\epsilon}^{-1}\epsilon_{\alpha\beta}.$
Then, expanding the friction force would yield
\begin{eqnarray}
\label{eqB16}
\left(\gamma_{\alpha\beta}^{-1}j_{\beta}\right)^{1} & =\left[\gamma_{0}^{-1}\delta_{\alpha\beta}-\frac{\psi}{2G}\phi_{0}\left(1-\phi_{0}\right)Q_{\alpha\beta}^{0}\gamma_{\epsilon}^{-1}\right]j_{\beta}^{1}+\gamma_{\epsilon}^{-1}j_{0}\epsilon_{\alpha x}^{1}.
\end{eqnarray}
 It is evident that the correction $\sim\epsilon_{\alpha x}^{1}$
appears in a similar way either due to $\gamma_{\epsilon}^{-1}$ or
$\nu^{1}$. Explicitly this yields the relation $\nu'=-\gamma_{\epsilon}^{-1}j_{0}.$
For $j_{0}<0$ and considering that the friction is expected to decrease
due to network alignment, we conclude that $\nu'$ is indeed expected
to be negative, as was mentioned above.

We return to the calculation of the divergence of the relative current
and examine each contribution separately. For the orientational field,
we find that
\begin{eqnarray}
\label{eqB17}
iq_{\alpha}h_{\alpha}^{1} & =-K\left(q^{2}p_{d}^{1}+l_{p}^{-1}q^{2}\phi^{1}\right)+h_{\parallel}^{0}p_{d}^{1}-2\phi_{0}\left(1-\phi_{0}\right)\psi iq_{\beta}\left(\epsilon_{\alpha\beta}p_{\beta}\right)^{1}.
\end{eqnarray}
The last contribution appears also in the $\nu'$ term. Summing the
two contributions leads to $j_{u}\left(u_{p}+\frac{1}{3}\phi_{0}^{2}\left(1-\phi\right)^{2}\frac{\psi}{G}\right)p_{d}^{1}$
, where $j_{u}/\gamma=-\nu'+2\lambda\phi_{0}\left(1-\phi_{0}\right)\psi$
describes the two contributions to the strain-dependent relative forces:
the active relative force $\sim\nu'$ and the permeation-alignment
mechanism $\sim\lambda,$ which includes polarization alignment to
the strain. We further simplify using the steady-state equation for
the current,
\begin{eqnarray}
\label{eqB18}
\frac{1}{3}\phi_{0}^{2}\left(1-\phi\right)^{2}\frac{\psi}{G}j_{u}+\gamma\phi_{0}\left(1-\phi_{0}\right)\left(\lambda h_{\parallel}+\nu\right) & =j_{0}.
\end{eqnarray}

For the matrix force, we calculate separately the contributions of
the elastic stress and osmotic pressure. The elastic contribution
is
\begin{eqnarray}
\label{eqB19}
\fl -q_{\alpha}q_{\beta}\sigma_{\alpha\beta}^{el,1} & =-q_{\alpha}q_{\beta}\left[2G\tilde{\epsilon}_{\alpha\beta}^{1}+B\epsilon^{1}\delta_{\alpha\beta}+2\phi_{0}\left(1-\phi_{0}\right)\psi p_{\alpha}^{1}p_{\beta}^{0}+\left(1-2\phi_{0}\right)\psi Q_{\alpha\beta}^{0}\phi^{1}+2\frac{\phi^{1}}{\phi_{0}}G\tilde{\epsilon}_{\alpha\beta}^{0}\right]\nonumber \\
 \fl & =\left[\frac{1}{\phi_{0}}\left(\frac{4}{3}G+B\right)-\frac{1}{3}\phi_{0}\psi\right]q^{2}\phi^{1},
\end{eqnarray}
where we have made use of the fact that $\epsilon^{1}=-\phi^{1}/\phi_{0}$
and $q_{x}=0.$ The contribution from the osmotic pressure is
\begin{eqnarray}
\label{eqB20}
\fl \phi_{0}q^{2}\overline{\mu}^{1} & =\phi_{0}q^{2}\left[\frac{k_{B}T}{a^{3}}\frac{\phi^{1}}{1-\phi_{0}}-2\left(\chi_{0}+\psi Q_{\alpha\beta}^{0}\epsilon_{\alpha\beta}^{0}\right)\phi^{1}+\left(1-2\phi_{0}\right)\psi\left(Q_{\alpha\beta}^{0}\epsilon_{\alpha\beta}^{1}+2p_{\alpha}^{1}p_{\beta}^{0}\epsilon_{\alpha\beta}^{0}\right)\right.\nonumber\\
\fl&\left.+\frac{2}{\phi_{0}}G\tilde{\epsilon}_{\alpha\beta}^{1}\tilde{\epsilon}_{\alpha\beta}^{0}+Kl_{p}^{-1}p_{d}^{1}+2\kappa q^{2}\phi^{1}\right]\nonumber\\
\fl & =\phi_{0}q^{2}\left[\frac{k_{B}T}{a^{3}}\frac{1}{1-\phi_{0}}-2\chi_{0}+\frac{2\psi^{2}}{3G}\phi_{0}\left(1-\phi_{0}\right)-\frac{1}{3}\psi+2\kappa q^{2}\right]\phi^{1}+\phi_{0}Kl_{p}^{-1}q^{2}p_{d}^{1}.
\end{eqnarray}
Together this yields
\begin{eqnarray}
\label{eqB21}
\fl iq_{\alpha}f_{\alpha}^{m1} & =\left[\frac{1}{\phi_{0}}\left(\frac{4}{3}G\left(1+\tau s\right)+B\left(1+\overline{\tau}s\right)\right)+\phi_{0}\chi^{-1}\left(1+l_{\phi}^{2}q^{2}\right)\right]q^{2}\phi^{1}+\phi_{0}Kl_{p}^{-1}q^{2}p_{d}^{1},
\end{eqnarray}
where we have defined the effective inverse susceptibility
\begin{eqnarray}
\label{eqB22}
\chi^{-1} & =\frac{k_{B}T}{a^{3}}\frac{1}{1-\phi_{0}}-2\left(\chi_{0}+\frac{\psi}{3}\right)+\frac{2\psi^{2}}{3G}\phi_{0}\left(1-\phi_{0}\right),
\end{eqnarray}
and the interfacial correlation length $l_{\phi}=\sqrt{2\kappa\chi}$.
Note that parallel cell-strain alignment ($\psi<0$) has a positive
contribution.

For the cellular force, we find that
\begin{eqnarray}
\label{eqB23}
iq_{\alpha}f_{\alpha}^{c1} & =\left(\overline{\zeta}-\frac{1}{3}\zeta\right)q^{2}\phi^{1}.
\end{eqnarray}
Note that the isotropic $\overline{\zeta}$ stress does not affect
the total stress, due to incompressibility. It simply renormalizes
the pressure $\delta P.$ Its only role is in the equation for the
relative current [Eq.~(\ref{eqB15})] and it
can be interpreted as an active, relative force $\sim\partial_{\alpha}\phi.$

Inserting back in the equation for the current yields overall
\begin{eqnarray}
\label{eqB24}
iq_{\alpha}j_{\alpha}^{1} & =\left(D_{\phi}+l_{\eta}^{2}s\right)q^{2}\phi^{1}+\left[\left(l_{p}^{-1}-\lambda\right)l_{\gamma1}^{2}D_{p}q^{2}+j_{0}+j_{u}u_{p}\right]p_{d}^{1}.
\end{eqnarray}
 Here we have defined the effective osmotic diffusion constant as
$D_{\phi}=D_{1}+D_{2}l_{\phi}^{2}q^{2}-\lambda l_{p}^{-1}l_{\gamma1}^{2}D_{p}$
with
\begin{eqnarray}
\label{eqB25}
D_{1} & =\gamma\phi_{0}\left(1-\phi_{0}\right)\chi^{-1}+\gamma\frac{1-\phi_{0}}{\phi_{0}}\left(\frac{4}{3}G+B\right)+\gamma\phi_{0}\left(\frac{1}{3}\zeta-\overline{\zeta}\right),\nonumber\\
D_{2} & =\gamma\phi_{0}\left(1-\phi_{0}\right)\chi^{-1},
\end{eqnarray}
as well as $l_{\eta}=\sqrt{\gamma\frac{1-\phi_{0}}{\phi_{0}}\left(\frac{4}{3}G\tau+B\overline{\tau}\right)}$,
a screening length due to the interplay between friction and transient
matrix viscosity. This diffusion constant differs from that in the isotropic case [Eq.~(\ref{eq10})] in two ways: its inverse susceptibility has contributions $\sim\psi$ [Eq.~(\ref{eqB22}], and it includes the nematic active stress $\sim\zeta$.

Inserting Eq.~(\ref{eqB24}) in Eqs. (\ref{eqB8}) and (\ref{eqB14}) yields the linearized dynamic equations, Eqs. (\ref{eq8}) and (\ref{eq9}).

\section{Linear stability in the rigid limit}
\label{AppC}
In the rigid matrix limit, concentration fluctuations generate a large
free-energetic cost, and $D_{\phi}$ becomes very large. For a finite
retardation time, $l_{\eta}^{2}$ becomes very large as well. We consider
a finite system size $L$ and a minimal wave vector $q_{m}=2\pi/L,$ such
that $l_{\eta}^{2}q_{m}^{2}\gg1$ and $D_{\phi}q_{m}^{2}\gg k_{\phi}.$
In this approximation, Eqs.~(\ref{eq8}) and (\ref{eq9}) reduce to
\begin{eqnarray}
\label{eqC1}
\fl 0 & =\left(D_{\phi}+l_{\eta}^{2}s\right)q^{2}\phi^{1}+\left[\left(l_{p}^{-1}-\lambda\right)l_{\gamma1}^{2}D_{p}q^{2}+j_{0}+j_{u}u_{p}\right]p_{d}^{1}\nonumber\\
\fl 0 & =\left[s+\left(\bar{\psi}-\lambda j_{u}\right)u_{p}+\left(1+\lambda\left(\lambda-l_{p}^{-1}\right)l_{\gamma1}^{2}\right)D_{p}q^{2}\right]p_{d}^{1}-\lambda\left(D_{\phi}+l_{\eta}^{2}s\right)q^{2}\phi^{1}.
\end{eqnarray}
One solution is $s=-D_{\phi}/l_{\eta}^{2},$which corresponds to stable
concentration fluctuations. The other two possible solutions are found
from the remaining factor in the determinant
\begin{eqnarray}
\label{eqC2}
\fl 0 & =s+\left(\bar{\psi}-\lambda j_{u}\right)u_{p}+\left(1+\lambda\left(\lambda-l_{p}^{-1}\right)l_{\gamma1}^{2}\right)D_{p}q^{2}+\lambda\left[\left(l_{p}^{-1}-\lambda\right)l_{\gamma1}^{2}D_{p}q^{2}+j_{0}+j_{u}u_{p}\right]\nonumber\\
\fl & =s+\bar{\psi} u_{p}+\lambda j_{0}+D_{p}q^{2}.
\end{eqnarray}
Substituting $u_{p}$ [Eq.~(\ref{eqB13})]
leads to Eq.~(\ref{eq11}).

\section{Estimations of parameters}
\label{AppD}
The basic time scale of the theory is $1/k_{\phi}.$ We estimate it
as $1/k_{\phi}=24$h for a typical division time of one day.
The basic length scale of the theory is the correlation length $l_{\phi}.$
For simplicity, we choose a small length of order of the cell size
$a$ that we set as $l_{\phi}=a=10\,\mu\mathrm{m}.$ This is the lowest value
that we consider for length scales, including $\lambda^{-1},$ $l_{p},$
and $\xi.$ Next, we estimate the remaining parameters of our theory. The estimations are summarized in Table~\ref{table2}.

\textbf{Osmotic diffusion constant.} The diffusion constant $D_{1}$
includes terms of the form $\gamma G,$ $\gamma\zeta,$ $\gamma\chi^{-1}$
[Eq.~(\ref{eqB25})]. The mobility can
be related to the cellular shear viscosity $\eta$ as $\gamma\approx\xi^{2}/\eta,$
where $\xi$ is a typical mesh size. The viscosity of epithelial monolayers
is of order $\eta\approx10^{3}-10^{4}\mathrm{Pa\,h}$~\cite{Blanch-Mercader2017}. As our theory coarse grains the cells and solvent together,
we consider the value of $\eta=1\,$kPa\,h. This value can be regarded as an upper bound of the viscosity. The ECM and collagen gels in general can have a large range
of stiffness values in the range $0.1<G<10\,\mathrm{kPa}$~\cite{Levental2007,Ray2018}.
For the active stress, we consider a 2D myosin contractility of $\zeta_{2D}\approx1\mathrm{kPa}\,\mu m$~\cite{Marchetti2013}.
Dividing by a typical cell size of $a=10\,\mu m,$ the cells are expected
to exert a stress of order $0.1$ $\mathrm{kPa}.$ We use this order of magnitude as well for extensile active stresses. For the inverse
susceptibility, we make a scaling argument, taking $1/k_{\phi}$ as
the basic timescale of the system. We write the corresponding term
in the diffusion constant as $D_{2}=\gamma\phi_{0}\left(1-\phi_{0}\right)\chi^{-1}\equiv l^{2}k_{\phi},$
where $l$ is a lengthscale. The minimal possible $l$ is $l=a.$
For a fixed $\chi,$ this is obtained for the minimal mobility $\gamma=a^{2}/\eta.$
This yields $D_{2}\approx\xi^{2}k_{\phi}$ and, consequently, $\chi^{-1}\approx\eta k_{\phi}\approx0.1$ kPa.

\textbf{Relative current and strain-polarization coupling. }The steady-state
relative current is estimated by a typical migration velocity~\cite{Ray2018}
$j_{0}=5\mu m/h$. We also consider the contribution of the strain-induced
current around the steady state, $\left(j_{u}+\lambda l_{\gamma1}^{2}\bar{\psi}\right)u_{p}\left(0\right).$
We have $j_{u}+\lambda l_{\gamma1}^{2}\bar{\psi}=-\gamma\nu'$.
The strain-dependent, active, relative force is estimated as $\nu'=-\gamma_{\epsilon}^{-1}j_{0},$
assuming that it has a similar effect as a strain-dependent friction
coefficient~(\ref{AppB}). For simplicity,
we consider $\gamma_{\epsilon}=-\gamma.$ This yields $j_{u}+\lambda l_{\gamma1}^{2}\bar{\psi}=-j_{0}.$
The polarization-induced strain parameter is given by $u_{p}\left(0\right)=\phi_{0}\left(1-\phi_{0}\right)\left[\left(1-\phi_{0}\right)\zeta+\gamma_{1}\lambda j_{0}\right]/\left(2G\right).$ For a small modulus $G=0.1$kPa, we
find that $u_{p}\left(0\right)\approx0.1\left(1-5\lambda l_{\phi}\right).$
The strain $u_{p}\left(0\right)$ often appears next to the strain-polarization
rate $\bar{\psi}=-2\psi/\gamma_{1}.$
For the strain-polarization coupling $\psi,$ we consider the value
$\phi_{0}\left(1-\phi_{0}\right)\psi=-0.1G.$ The sign signifies that
the cells align parallel to network segments and the order of magnitude
is the largest possible within the framework of linear elasticity.
The product $\bar{\psi} u_{p}\left(0\right)$ is then given by $0.25\left(1-5\lambda l_{\phi}\right)k_{\phi}$,
where we have set the rotational viscosity as the shear viscosity,
$\gamma_{1}=\eta.$

\textbf{Angular diffusion constant. }The angular diffusion constant
is $D_{p}=K/\gamma_{1}.$ The Frank constant in two dimensions $K_{2D}$
can be estimated from experiments that measure the active lengthscale,
$l_{a}=2\pi\sqrt{K_{2D}/\left|\zeta_{2D}\right|}$, where $\zeta_{2D}$
is the two-dimensional active, nematic stress. Experiments on cell
monolayers have measured a length of order $l_{\alpha}\approx50\,\mu m$
~\cite{Duclos2018}. Considering again a 2D myosin contractility of order kPa~$\mu$m,
we find that $K_{2D}\approx100\,\mathrm{kPa}\,\mu \mathrm{m}^3.$ Dividing
by the cell size yields $K_{3D}\approx10\,\mathrm{kPa}\,\mu \mathrm{m}^2$ and $D_{p}\approx10\,\mu \mathrm{m}^{2}/\mathrm{h}$.

\begin{table}[ht]
	\caption{Estimations of the parameters used in our theory. Ranges of values result from the range of elastic moduli, $0.1<G<10$\,kPa. The screening length $l_\eta$ is evaluated for large elastic moduli, as in Sec.~\ref{sec6}.}
	\begin{adjustbox}{width=1\textwidth}
		\begin{tabular}{| c | l |c | l |}
			\mr
			\bf Parameter & \bf Estimate & \bf Parameter & \bf Estimate \\ \hline
			$l_\phi$ & $10\,\mu$m & $\xi$ (mesh size) & $l_\phi<\xi<10l_\phi$\\\hline
			$D_1\cdot\left(l_\phi/\xi\right)^2$ & $-10<...<10^3\,\mu$m$^2$/h & $D_2\cdot\left(l_\phi/\xi\right)^2$ & $5\,\mu$m$^2$/h \\ \hline
						$k_\phi$ & $1/24$h & $D_p$ &$10\,\mu$m$^2$/h   \\ \hline
			$l_\eta$ & $l_{\eta}\approx\xi$ & $l_{\gamma1}$ & $l_{\gamma1}\approx\xi$\\ \hline
			$\bar{\psi}$	& $10^{-2}<...<1$/h   & $u_p$ & $\bar{\psi} u_p(0)\approx 10^{-2}\left(1-5\lambda l_\phi\right)$/h \\ \hline
			$j_0$ & $-5\,\mu$m/h & $j_u$ & $j_u\approx-j_0-\lambda\xi^2\bar{\psi} $ \\ \hline
			$l_p$ & $|l_p|>l_\phi$ & $\lambda$ & $0>\lambda>-1/l_\phi$\\ \hline
			
		\end{tabular}
	\end{adjustbox}
	\label{tableD}
\end{table}

\clearpage
\bibliographystyle{unsrt}
\bibliography{refs}
\end{document}